\documentclass[12pt]{article}
\usepackage{graphicx}
\usepackage{amssymb}
\usepackage{amsmath}
\usepackage{bm}
\usepackage{cite}
    
\newcommand{\bear}{\begin{array}}  
\newcommand {\eear}{\end{array}}
\newcommand{\bea}{\begin{eqnarray}}   
\newcommand{\eea}{\end{eqnarray}}
\newcommand{\beq}{\begin{eqnarray}}   
\newcommand{\eeq}{\end{eqnarray}}
\newcommand{\bef}{\begin{figure}}  
\newcommand {\eef}{\end{figure}}
\newcommand{\bec}{\begin{center}}  
\newcommand {\eec}{\end{center}}

\newcommand{\oneui}{\overline{\tilde{\chi}^0}}
\newcommand{\sla}[1]{\not\!\! #1}

\begin{document}

\begin{titlepage}

\begin{flushright}
IPMU~12-0187 \\
CALT 68-2891 \\
\end{flushright}

\vskip 1.35cm
\begin{center}

{\large 
{\bf 
Direct Search of Dark Matter in High-Scale Supersymmetry
}
}

\vskip 1.2cm

Junji Hisano$^{a,b}$,
Koji Ishiwata$^c$
and
Natsumi Nagata$^{a,d}$ \\

\vskip 0.4cm

{ \it $^a$Department of Physics, 
Nagoya University, Nagoya 464-8602, Japan}\\
{\it $^b$Kavli Institute for the Physics and Mathematics of the Universe
 (Kavli IPMU), University of Tokyo, Kashiwa 277-8584, Japan}\\
{\it $^c$California Institute of Technology, Pasadena, CA 91125, USA}\\
{\it $^d$Department of Physics, 
University of Tokyo, Tokyo 113-0033, Japan}

\date{\today}

\begin{abstract} 
 We study direct detection of dark matter in a supersymmetric (SUSY)
 model where most SUSY particles have very high-scale masses
 beyond the weak scale. In the scenario, a Wino-like or a Higgsino-like
 neutralino is a good candidate for the dark matter in the
 Universe. The neutralino scatters off nuclei
 by a Higgs boson exchange diagram and  also electroweak loop diagrams.
 It is found that the elastic-scattering cross section with nuclei is
 enhanced or suppressed due to constructive or deconstructive
 interference among the diagrams. Such a cross section is within the
 reach of future experiment in some parameter region.

\end{abstract}

\end{center}
\end{titlepage}

\section{Introduction}
\label{sec:intro}

Supersymmetric (SUSY) remodeling of the Standard Model (SM) is one of
the promising candidates for physics beyond the SM. The minimal
extension, called the minimal supersymmetric Standard Model (MSSM),
has been studied enthusiastically in various literature. The
weak-scale SUSY is, however, severely constrained by the experiments
at the Large Hadron Collider (LHC). Since no signal of SUSY particles
has been discovered yet, the ATLAS and the CMS Collaborations have
imposed stringent limits on their masses, especially those of
colored particles \cite{:2012rz}. The weak-scale SUSY is also
challenged by the discovery of the SM-like Higgs boson with a
mass of about 125~GeV, which is recently reported by the
collaborations \cite{:2012gk}. In the MSSM radiative corrections from
heavy sfermions make the Higgs mass larger \cite{Okada:1990vk}. Thus
those results from the LHC may indicate that the SUSY scale is
somewhat higher than the weak scale \cite{Hall:2011aa}.

Although the high-scale SUSY scenario sounds unnatural in a
viewpoint of the hierarchy problem, phenomenological aspects of heavy
sfermions are quite fascinating \cite{Wells:2003tf,
  ArkaniHamed:2004fb, Giudice:2004tc, ArkaniHamed:2004yi,
  Wells:2004di, Hall:2009nd, Hall:2011jd}.  Because of the sufficient
radiative corrections, the 125~GeV Higgs boson may be achieved
\cite{Giudice:2011cg}. The SUSY contributions to flavor changing
neutral current (FCNC) processes and electric dipole moments
are suppressed by heavy sfermion masses so that the SUSY
flavor and CP problems are relaxed \cite{Gabbiani:1996hi}. In
cosmology, the gravitino problem may be avoided because it may be as
heavy as sfermions, then the thermal leptogenesis for baryon asymmetry
in the Universe works with high reheating temperature
\cite{Fukugita:1986hr}. On top of that, the gauge coupling unification
is achieved as precisely as that in the MSSM since the sfermions form
the SU(5) multiplets, and the proton lifetime could be well above the
current experimental limit \cite{Hisano:2012wq}.  These features have
stimulated various works~\cite{Jeong:2011sg}.

The high-scale SUSY scenario does not necessarily mean all SUSY
particles in the MSSM are heavy. Superpartners of gauge bosons and Higgs
bosons may be at the weak scale without destroying the above
features. This is plausible because the lightest particle among their
mixed states is the lightest SUSY particle (LSP) and it is a good
candidate for the dark matter (DM) in the Universe.  Such a candidate is
one of the so-called
Weakly Interacting Massive particles (WIMPs).  Though sfermions may be
beyond the reach of the LHC, the LSP DM may be searched in the direct
dark matter detection experiments.

In this article we consider a scenario where the LSP mass is around
the weak scale and the other SUSY particles are much heavier, and we give
a precise calculation of an elastic-scattering cross section between the
LSP DM and nucleon.  A Wino-like or Higgsino-like neutralino is
a viable DM candidate since the thermal production in the early hot universe
 gives the observed DM density even in the
heavy sfermion scenario; Wino with a mass of 2.7--3.0~TeV
\cite{Hisano:2006nn} or Higgsino with a mass of 1~TeV
\cite{Cirelli:2007xd}. The neutralino mass less than TeV is also possible to
explain the DM density when its non-thermal production is
considered \cite{Gherghetta:1999sw, Moroi:1999zb}. Also the Wino LSP
is a natural consequence of the anomaly mediation
\cite{Randall:1998uk}.  Therefore we focus on those well-motivated
cases.  Since there are only a few undetermined parameters, the
observed value of the Higgs boson mass at 125~GeV allows us to make a
robust prediction for the scattering cross section with nucleon.  As we
will see below, both the tree-level \cite{Murakami:2000me, Moroi:2011ab}
and the loop-level processes \cite{Hisano:2010fy, Hisano:2010ct,
  Hisano:2011cs} give rise to sizable contributions to the scattering
cross section.

This paper is organized as follows. In the next section, we explain
the scenario of high-scale supersymmetry, in which the Wino-like or
Higgsino-like neutralino is predicted as the DM. In Sec.~3, the effective
Lagrangian for the neutralino-nucleon elastic scattering is
reviewed. In Sec.~4, we discuss relevant tree- and loop-level
contributions to the spin-independent (SI) scattering of the Wino-like or
Higgsino-like neutralino, and evaluate the cross section. Section 5 is devoted
to the conclusion.  In the evaluation of the SI cross section in the text, we
use the results of the lattice QCD simulations for the mass fractions of
light quarks in the nucleon.  We also show the results when the
mass fractions estimated in the chiral perturbation theory are used in
Appendix A.
In Appendix B, we give the spin-dependent (SD) cross section for
completeness.

\section{The Scenario}
\label{sec:model}

In this section we briefly describe the scenario that we discuss in
this paper. As it is mentioned in the Introduction, we consider a SUSY
scenario where all SUSY particles are well above the weak scale,
except for Wino or Higgsino.  Such mass spectrum is given by a simple
SUSY breaking mechanism
~\cite{Wells:2003tf,ArkaniHamed:2004fb,Randall:1998uk,Hall:2011jd}. Assume
that there exists a SUSY breaking hidden sector containing a
SUSY breaking field $Z$ which is charged under some symmetry. Then a
generic form of K\"{a}hler potential yields 
masses of $M_{\rm SUSY}\sim F_Z/M_*$ for all the scalar bosons in the
MSSM except the lightest Higgs boson ($F_Z$ and $M_*$ are the
$F$-component vacuum expectation value (VEV) of the field $Z$ and the
messenger scale, respectively). 
On the other hand, since $Z$ is charged under some symmetry, the gaugino
and Higgsino mass terms are not given by the $Z$-field linear
terms. Thus they do not necessarily have the mass scale $M_{\rm
  SUSY}$, {\it i.e.}, they are model dependent.\footnote{
Trilinear soft SUSY breaking terms are also model dependent. However,
  they are irrelevant in our discussion.
}
Now we consider the case of $M_*=M_{\rm Pl}$
($M_{\rm Pl}$ is the reduced Planck scale). In the case,
 gauginos acquire their masses via the anomaly mediation, which are
of the order of $m_{3/2}/16\pi^2$. Here $m_{3/2}=F_Z/\sqrt{3}M_{\rm
  Pl}$ is the gravitino
mass. For Higgsino, on the other hand, the so-called $\mu$ term,
 $\mu H_u H_d$ in the superpotential ($H_u$ and $H_d$ are up-type and
 down-type Higgs chiral superfields, respectively), may be absent
by a certain symmetry, {\it e.g.}, the Peccei-Quinn symmetry
\cite{Peccei:1977hh}.  In such a case, the gaugino-Higgs loops induce
the Higgsino mass, which is smaller than the gaugino masses by another
loop factor. Thus, the Higgsino-like neutralino is the LSP. On the
contrary it may be as heavy as a gravitino in another case. When the
K\"{a}hler potential has a term, $K=\kappa H_uH_d+\cdots$, the
Higgsino mass is provided by the supergravity effects and it lies
around the gravitino mass scale.  In that case, the Wino-like
neutralino is the LSP.

In order to consider the Wino-like or Higgsino-like neutralino,
we take the Wino and Higgsino mass parameters as free
parameters. Namely, the gaugino and Higgsino mass terms are given by
\begin{eqnarray}
{\cal L}^{M}_{\rm ino}=
- \sum_{a=1,2,3}\frac{1}{2}M_{a}\tilde{\lambda}_a \tilde{\lambda}_a 
- \mu \tilde{H}_u \tilde{H}_d,
\label{eq:Lm_ino}
\end{eqnarray}
where $\tilde{H}_u \tilde{H}_d = \tilde{H}_u^+ \tilde{H}_d^--\tilde{H}_u^0
\tilde{H}_d^0$ and
\begin{eqnarray}
 M_{a}=\frac{b_a g_a^2}{16\pi^2} m_{3/2}.
\end{eqnarray}
Here $\tilde{\lambda}_a(a=1,2$ and 3) are Bino, Wino and gluino,
respectively, and $\tilde{H}_u$ and $\tilde{H}_d$ are 
up-type and down-type Higgsinos, respectively. 
The coefficients $b_a$ denote the one-loop beta functions of
the gauge coupling constants $g_a$ ($a=1,2$, and $3$ for U(1)$_Y$,
SU(2)$_L$, and SU(3)$_C$, respectively), given by
$(b_1,b_2,b_3)=(33/5,1,-3)$.\footnote
{Threshold correction may change the ratio $M_1/M_2$ from the above
  relation. However, this does not affect our numerical result
  significantly.}
After the electroweak symmetry breaking, Bino ($\tilde{B}$), Wino
($\tilde{W}^0$), and neutral Higgsinos mix with each other. The mass
eigenstates, called neutralinos, are obtained as $\tilde{\chi}^0_i =
\sum_j Z_{ij}\chi^0_j$, where
$\chi^0_i=\tilde{B},\tilde{W}^0,\tilde{H}^0_u$, and $\tilde{H}^0_d$ for
$i=1,2,3$, and $4$, respectively. $\tilde{\chi}^0_1$ is the lightest
neutralino and from now on we omit the
subscript of $\tilde{\chi}^0_1$ for simplicity. In the following
calculation we take $M_2$ as a free parameter instead of $m_{3/2}$,
and consider the case where the lightest neutralino explains the
current relic density of DM. 

When sfermions are very heavy, the 125~GeV SM-like Higgs boson is
achieved with $\tan \beta\sim 1$--$5$~\cite{Giudice:2011cg}. Here
$\tan \beta$ is the ratio of the VEVs of up- and down-type Higgs
fields. In this article, however, we also consider larger $\tan \beta$
to demonstrate the cross section for a general case, assuming
appropriate sfermion masses to make the Higgs boson mass 125~GeV.

\section{Neutralino-Nucleon Scattering Cross Section}
\label{sec:eff}

Here we give formulae for the calculation of the scattering cross
section of the neutralino with nucleon \cite{Hisano:2010ct, Drees:1993bu,
Jungman:1995df}. It is calculated from the effective
Lagrangian for scattering of the neutralino with quarks and gluon in
the limit of low relative velocity, which is given by
\begin{eqnarray} 
{\cal L}_{\rm eff}&=&  d_q\
  \oneui\gamma^{\mu}\gamma_5\tilde{\chi}^0\
  \bar{q}\gamma_{\mu}\gamma_5 q + f_q m_q\ \oneui\tilde{\chi}^0\
  \bar{q}q \cr &+& \frac{g^{(1)}_q}{M} \ \oneui
  i \partial^{\mu}\gamma^{\nu} \tilde{\chi}^0 \ {\cal O}_{\mu\nu}^q +
  \frac{g^{(2)}_q}{M^2}\ \oneui(i \partial^{\mu})(i \partial^{\nu})
  \tilde{\chi}^0 \ {\cal O}_{\mu\nu}^q \nonumber \\ &+& 
{f}_G\ 
\oneui\tilde{\chi}^0 G_{\mu\nu}^aG^{a\mu\nu}
+\frac{g^{(1)}_G}{M}\
 \overline{\tilde{\chi}^0} i \partial^{\mu}\gamma^{\nu}
\tilde{\chi}^0 \ {\cal O}_{\mu\nu}^g
+
\frac{g^{(2)}_G}{M^2}\
\overline{\tilde{\chi}^0}(i\partial^{\mu}) (i\partial^{\nu})\tilde{\chi}^0 
\
{\cal O}_{\mu\nu}^g \ ,
\label{eff_la}
\end{eqnarray}
where $M$ and $m_q$ are the masses of the neutralino and quarks,
respectively. Sum over quark flavors $q=u,d,s$ for the first and second
terms and $q=u,d,s, c, b$ for the third and fourth terms is implicit.
The field strength tensor of the gluon field is denoted by $G^a_{\mu
\nu}$. The last two lines include the quark and gluon twist-2 operators,
${\cal O}_{\mu\nu}^q$ and ${\cal O}_{\mu\nu}^g$, respectively, which
are defined as,
\beq {\cal O}_{\mu\nu}^q&\equiv&\frac12 \bar{q} i
\left[D_{\mu}\gamma_{\nu} +
  D_{\nu}\gamma_{\mu} -(g_{\mu\nu}/2)\! \sla{D}
\right] q \ ,
\nonumber\\
{\cal O}_{\mu\nu}^g&\equiv&G_{\mu}^{a\rho}G_{\rho\nu}^{a}+
  (g_{\mu\nu}/4) G^a_{\alpha\beta}G^{a\alpha\beta} \ ,
\label{twist2}
\eeq
with $D_\mu\equiv\partial_\mu-i g_3A^a_\mu T_a$ the covariant
derivative ($T_a$ is the generator of SU(3)$_C$, and $A^a_\mu$ is the
gluon field). In order to remove the redundant terms, we use the
integration by parts and the equation of motion for the operators. The
first term in Eq.~(\ref{eff_la}) yields the SD interaction, while the
other terms generate the SI interactions. 

In order to compute the $\tilde{\chi}^0$-nucleon cross section from
the effective Lagrangian, we need to evaluate the nucleon matrix elements
of the quark and gluon operators.  The nucleon matrix elements of the
scalar-type light-quark operators, {\it i.e.}, $m_q\bar{q}q$
($q=u,d,s$), are parametrized as
\begin{equation}
\langle N \vert m_q \bar{q} q \vert N\rangle
\equiv m_N f_{Tq}^{(N)}\ , 
\label{ftq}
\end{equation}
where $m_N$ is the nucleon mass and $\vert N\rangle$ denotes the
one-particle state of the nucleon $(N=p,n)$. For the heavy quarks and
gluon, on the other hand, their matrix elements are obtained by using
the trace anomaly of the energy-momentum tensor in QCD:
\begin{align}
\langle N \vert m_Q \bar{Q} Q \vert N\rangle  &=-\frac{\alpha_s}{12\pi}
c_Q \langle N\vert G^a_{\mu\nu}G^{a\mu\nu}\vert N\rangle~,
\nonumber \\
m_Nf^{(N)}_{TG}&=-\frac{9\alpha_s}{8\pi}
\langle N\vert G^a_{\mu\nu}G^{a\mu\nu}\vert N\rangle ~,
\label{mat_QG}
\end{align}
with $f^{(N)}_{TG}\equiv 1-\sum_{q}f^{(N)}_{Tq}$ and $\alpha_s\equiv
g_3^2/4\pi$. The long-distance QCD correction $c_Q$ in the above
expression is evaluated in Ref.~\cite{Djouadi2000} as
$c_Q=1+11\alpha_s(m_Q)/4\pi$, and we take their numerical values as
$c_c=1.32$, $c_b=1.19$, and $c_t=1$ in this paper. As can be seen
from Eq.~(\ref{mat_QG}), the scalar-type heavy quark operators
contribute to the nucleon matrix elements only through the loop-induced
gluon operator. 

The nucleon matrix elements of the twist-2 operators are evaluated with
the parton distribution functions (PDFs):
\begin{eqnarray}
&& \langle N(k)\vert 
{\cal O}_{\mu\nu}^q
\vert N(k) \rangle 
=\frac{1}{m_N}
(k_{\mu}k_{\nu}-m^2_N g_{\mu\nu}/4)\
(q_N(2)+\bar{q}_N(2)) \ ,
\label{nucleon_matrices} \\
&& 
\langle N(k) \vert 
{\cal O}_{\mu\nu}^g
\vert N(k) \rangle
= \frac{1}{m_N}
(k_{\mu}k_{\nu}-m^2_N g_{\mu\nu}/4)\ 
G_N(2) \ ,
\end{eqnarray}
where $q_N(2)$, $\bar{q}_N(2)$ and $G_N(2)$ are the second moments of
PDFs of quark, anti-quark and gluon, respectively, which are given by
\begin{eqnarray}
q_N(2)+ \bar{q}_N(2) &=&\int^{1}_{0} dx ~x~ [q_N(x)+\bar{q}_N(x)] \ ,
\cr
G_N(2) &=&\int^{1}_{0} dx ~x ~g_N(x) \ .
\end{eqnarray}
Here, we use the PDFs at the scale of $\mu=m_Z$ ($m_Z$ is the $Z$
boson mass), since, as will be described later, the terms with quark
twist-2 operators in Eq.~(\ref{eff_la}) are induced by the one-loop
diagrams in which the loop momentum around the weak boson mass scale
yields dominant contribution.

Finally, the SI effective coupling is obtained as
\begin{eqnarray}
 \frac{f_N}{m_N}=
 f_{Tq}^{(N)}f_q
+
\frac{3}{4} \left(q_N(2)+\bar{q}_N(2)\right)\left(g_q^{(1)}+g_q^{(2)}\right)
-\frac{8\pi}{9\alpha_s}f_{TG}^{(N)} f_G 
+\frac{3}{4} G_N(2)\left(g^{(1)}_G
+g^{(2)}_G\right)  
.
\label{f}
\end{eqnarray}
Here the sum of quark flavors is implicit as in
Eq.~(\ref{eff_la}).  Note the factor $1/ \alpha_s$ in front of $f_G$
in Eq.~(\ref{f}). It makes the gluon contribution sizable, although
the interactions of the neutralino with gluon are induced by
higher-loop processes than those with light quarks
\cite{Hisano:2010ct}. On the other hand, the contributions of the
twist-2 operators of gluon are subdominant\footnote{ Evaluating the
  nucleon matrix elements of twist-2 operators at $\mu=m_Z$ makes the
  perturbative expansion with respect to $\alpha_s$ reliable. Instead,
  if one would like to estimate the matrix elements at $\mu=1$~GeV,
  one also needs to take into account the operator-mixing effects due
  to the QCD radiative corrections, and therefore, to include the
  gluon twist-2 operator, as in Ref.~\cite{Hill:2011be}.  These two
  approaches are equivalent since the sum of the terms with twist-2
  operators is scale independent once they are multiplied by their
  coefficients. Generally speaking, however, the former approach makes
  the calculation robust thanks to the perturbativity of the QCD
  coupling.  } as $g^{(1)}_G$ and $g^{(2)}_G$ are suppressed by the
strong coupling constant $\alpha_s$. Thus, we ignore them in this
paper.

The effective axial vector coupling, which is relevant for the SD cross
section, is readily written as
\begin{eqnarray}
a_{N}&=& d_q \Delta q_N \ ,
\label{an}
\end{eqnarray}
with
\begin{equation}
 \langle N \vert 
\bar{q}\gamma_{\mu}\gamma_5  q \vert N \rangle = 2 s_{\mu}\Delta q_N \ .
\label{sfra}
\end{equation}
Here $s_{\mu}$ is the spin of the nucleon and the quark flavor sum is
taken for $q=u,d,s$.  By using the effective couplings
obtained above, we obtain the cross section of the neutralino with
nucleon:
\begin{eqnarray}
  \sigma_N&=&
  \frac{4}{\pi}m_R^2
  \left[\left| f_N\right|^2+3\left|a_N\right|^2\right]\ , 
\label{sigma}
\end{eqnarray}
where $m_R\equiv M m_N/(M+m_N)$ is the reduced mass of neutralino-nucleon
  system.

Before concluding this section, we refer to the numerical values for
the parameters that we use in this paper. The mass fractions of light
quarks, $f_{Tq}^{(N)}$ defined in Eq.~(\ref{ftq}), are to be extracted
from the results of the lattice QCD simulations \cite{Young:2009zb,
  :2012sa}. The mass fractions of light quarks are evaluated with
independent methods and also by independent groups so that the results
derived with the lattice QCD simulations have become more reliable. The mass
fractions evaluated in the chiral perturbation theory (ChPT) have larger
uncertainties than those from the lattice QCD. The SI cross section
evaluated by the use of the mass fractions from the ChPT, which predicts
larger $f_{Ts}^{(N)}$, is shown in  Appendix A for comparison. The second
moments of the PDFs of quarks and anti-quarks are calculated using the
CTEQ parton distribution \cite{Pumplin:2002vw}. The spin fractions,
$\Delta q_N$, in Eq.~(\ref{sfra}) are obtained from
Ref.~\cite{Adams:1995ufa}. In Table~\ref{table} we list the numerical
values for the mass fractions of both proton and neutron as well as the
second moments of the PDFs and the spin fractions for the proton. The second
moments and the spin fractions for the neutron are to be obtained by
exchanging the values of an up quark for those of a down quark.

\begin{table}
\begin{center}
\begin{tabular}{|l|l|}
\hline
\multicolumn{2}{|c|}{Mass fraction}\cr
\hline
\multicolumn{2}{|c|}{(proton) }\cr
\hline 
$f^{(p)}_{Tu}$& 0.019(5)\cr
$f^{(p)}_{Td}$& 0.027(6)\cr
$f^{(p)}_{Ts}$&0.009(22)\cr
\hline
\multicolumn{2}{|c|}{(neutron)}\cr
\hline
$f^{(n)}_{Tu}$&0.013(3)\cr
$f^{(n)}_{Td}$& 0.040(9)\cr
$f^{(n)}_{Ts}$& 0.009(22) \cr
\hline
\end{tabular}
\hskip 1cm 
\begin{tabular}{|l|l||l|l|}
\hline
\multicolumn{4}{|c|}{Second moment at $\mu=m_Z$ }\cr
\hline
$u(2)$&0.22&$\bar{u}(2)$& 0.034\cr
$d(2)$&0.11&$\bar{d}(2)$&0.036\cr
$s(2)$&0.026&$\bar{s}(2)$&0.026\cr
$c(2)$&0.019&$\bar{c}(2)$&0.019\cr
$b(2)$&0.012&$\bar{b}(2)$&0.012\cr
\hline
\end{tabular}
\hskip 1cm 
\begin{tabular}{|l|l|}
\hline
\multicolumn{2}{|c|}{Spin fraction }\cr
\hline
$\Delta u_p$& 0.77\cr
$\Delta d_p$& $-$0.49\cr
$\Delta s_p$& $-$0.15\cr
\hline 
\end{tabular}
\end{center}
\caption{Parameters for quark and gluon matrix elements. Errors are
 shown only for the mass fractions, which are used for comparison with
 the cross section evaluated with the mass fraction from the ChPT. } 
\label{table}
\end{table}
%

\section{Results}
\label{sec:results}

In this section we calculate the SI cross section for the Wino-like or
Higgsino-like neutralino in the high-scale SUSY scenario. The
values of the Higgsino and Wino mass parameters are model dependent.  Therefore, we
regard both $M_2$ and $\mu$ as free parameters in the following
analysis while we take $M_2$ positive.\footnote{
We assume the parameters to be real in this
article. Possible phases of the parameters might affect the ``Higgs''
contribution, which is defined later, to the SI effective coupling.}

\begin{figure}[t]
  \begin{center}
    \includegraphics[scale=0.5]{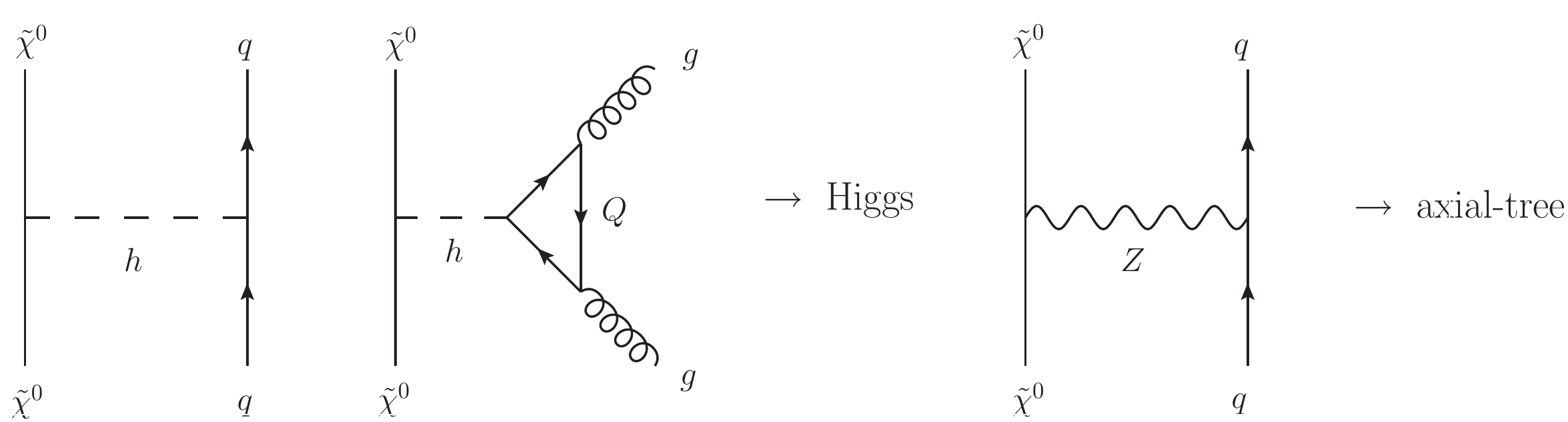}
  \end{center}
  \caption{Diagrams via tree-level
    $\tilde{\chi}^0$-$\tilde{\chi}^0$-Higgs/$Z$ interaction in
     elastic $\tilde{\chi}^0$-nucleon scattering. ``Higgs'' contribution and
    ``axial-tree'' contribution are defined in Eqs.~(\ref{def_fN}) and
    (\ref{def_aN}).}
  \label{fig:Higgs}
\end{figure}

\begin{figure}[t]
  \begin{center}
    \includegraphics[scale=0.5]{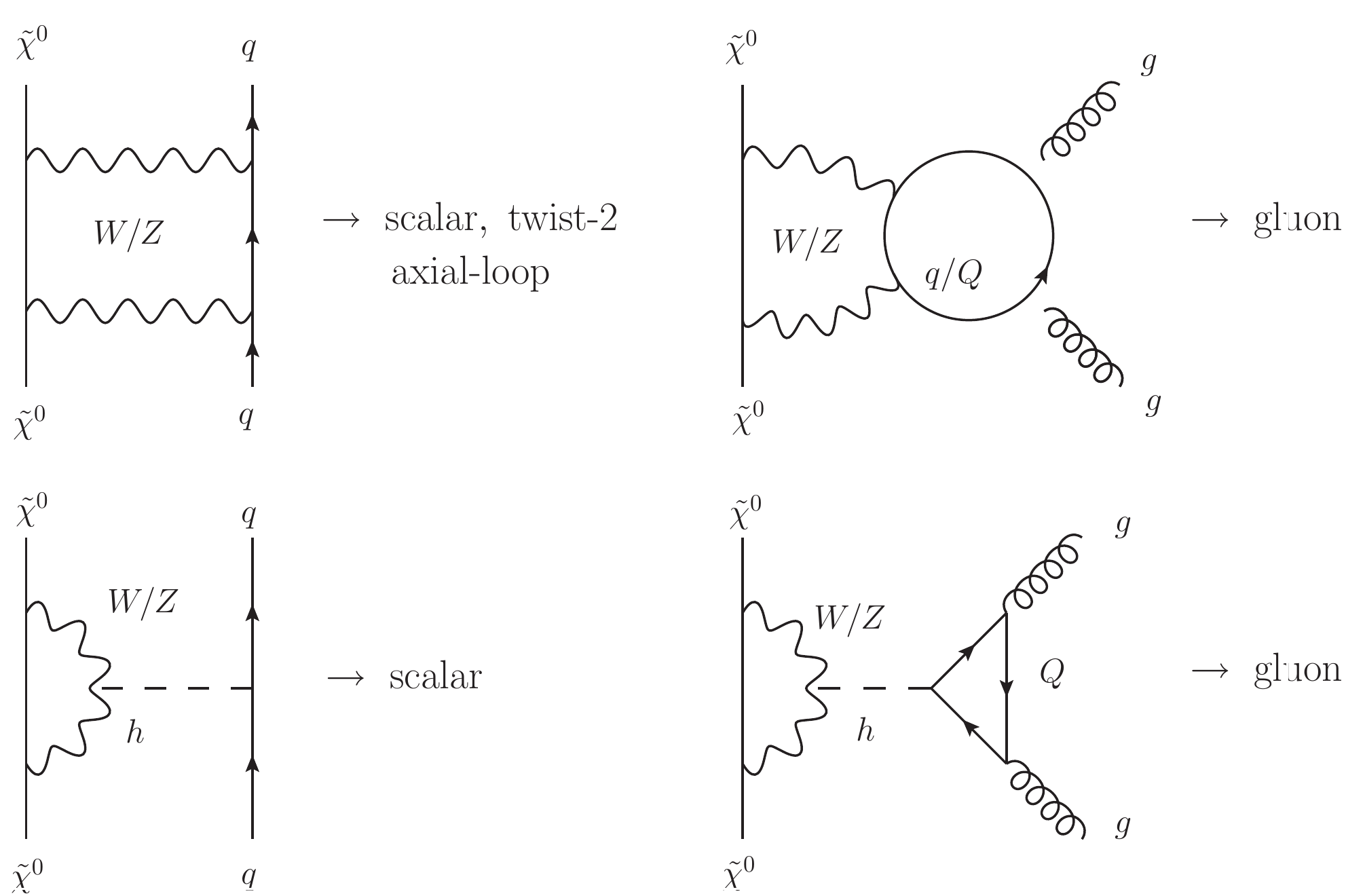}
  \end{center}
  \caption{Diagrams which are induced by electroweak interaction in
    elastic $\tilde{\chi}^0$-nucleon scattering.  ``scalar'',
    ``twist-2'', ``gluon'' and ``axial-loop'' correspond to each term
    in the effective couplings. Their definitions are given in
    Eqs.~(\ref{def_fN}) and (\ref{def_aN}). A complete set of diagrams
    is given in Ref.~\cite{Hisano:2011cs}.}
  \label{fig:Loop}
\end{figure}

The tree-level $\tilde{\chi}^0$-$\tilde{\chi}^0$-Higgs
  interaction yields scalar-type effective operators,
  $\oneui\tilde{\chi}^0\bar{q}q$ and $\oneui\tilde{\chi}^0
  G_{\mu\nu}^aG^{a\mu\nu}$ (Left in Fig.~\ref{fig:Higgs}).  Let us
  denote these contributions by $f_q^{H}$ and $f_G^{H}$, respectively.
  By adding them to the loop-level contributions\footnote{ 
Those contributions are evaluated in a pure Wino or Higgsino limit.
When $M_2\simeq\mu$, one needs to take the mixing among them
     into account and modify the formulae in Ref.~\cite{Hisano:2011cs}
     appropriately. In the present situation, however, the tree-level
     contributions dominate the loop-loop level ones. 
     Thus, the modification in the loop-level effects has no
     significance on the resultant scattering cross section. There
     exists a case where the tree-level contribution is still
     subdominant even when $M_2\simeq\mu$. As we will see later,
     however, the lightest neutralino is almost pure gauge eigenstate
     in such a case. Thus the results in Ref.~\cite{Hisano:2011cs} are
     applicable.} $f_q^{\rm EWIMP}$ and $f_G^{\rm EWIMP}$ which are
     induced via the $W/Z$ boson loop diagrams \cite{Hisano:2011cs}, we
     obtain
\begin{eqnarray}
f_q &=&  f_q^{H} +  f_q^{\rm EWIMP},
\\ 
f_G  &=& f_G^{H} +  f_G^{\rm EWIMP},
\end{eqnarray}
with
\begin{eqnarray}
f_q^{H} &=& \frac{g_2^2 s^h}{4 m_W m_h^2}, 
\\
f_G^{H} &=& -\frac{\alpha_s}{12 \pi} \sum_{Q=c,b,t}c_Q f_Q^{H},
\end{eqnarray}
where $m_W$ and $m_h$ are the masses for $W$ boson and Higgs boson,
respectively. The coupling of the neutralino with the Higgs
boson is denoted by $s^h$ in the above expression, which is given as 
\begin{eqnarray}
s^h = (Z_{12}-Z_{11}\tan \theta_W ) (Z_{13}\cos\beta -  Z_{14}\sin \beta).
\end{eqnarray}
Here $\theta_W$ is the weak mixing angle and we take the decoupling
limit since the heavier Higgs bosons have masses much larger than the
weak scale.  In addition, when $M_2$, $|\mu|\gtrsim m_W$, the coupling
$s^h$ is approximated as
\begin{eqnarray}
s^h \simeq 
\frac{m_W}{M_2^2-\mu^2}(M_2+\mu \sin 2 \beta),
\label{eq:Wino_sh}
\end{eqnarray}
in the Wino-like neutralino case and 
\begin{eqnarray}
s^h \simeq -\frac{1}{2}
\left[\frac{m_W}{M_2-|\mu|}+\frac{m_W\tan^2 \theta_W}{M_1-|\mu|}\right]
(1\pm \sin 2 \beta),
\label{eq:Higgsino_sh}
\end{eqnarray}
in the Higgsino-like neutralino case. Here the plus (minus) sign in front
of $\sin 2 \beta$ is for $\mu>0$ ($\mu<0$).\footnote
{There is an sign error in Eq.~(24) of Ref.~\cite{Hisano:2004pv} for
  $\mu<0$ case. In addition, for heavy Higgs coupling, the correct
  expression is $\pm \frac{1}{2}
  \left[\frac{m_W}{M_2-|\mu|}+\frac{m_W\tan^2
      \theta_W}{M_1-|\mu|}\right] \cos 2 \beta$.}

As it is seen in Eqs.~(\ref{eq:Wino_sh}) and
   (\ref{eq:Higgsino_sh}), in the case where one mass parameter is
much larger than the other ({\it i.e.}, $M_2\ll |\mu|$ or $|\mu|\ll M_2$), the
lightest neutralino becomes almost pure Wino or Higgsino state. 
Then the tree-level $\tilde{\chi}^0$-$\tilde{\chi}^0$-Higgs
interaction, as well as $\tilde{\chi}^0$-$\tilde{\chi}^0$-$Z$
interaction which is relevant for the SD scattering, is suppressed.
Thus the loop-level processes become important.  The loop-level
effective couplings are calculated in the previous
work~\cite{Hisano:2011cs}, where the elastic scattering cross section
for generic electroweak-interacting DM particles ({\it i.e.}, $n$-tuplet
of SU(2)$_L$ with hypercharge $Y$ of U(1)$_Y$) is evaluated. Pure Wino
corresponds to $n=3$ and $Y=0$, while pure Higgsino corresponds to
$n=2$ with $Y=1/2$. The previous results have revealed that the loop-level
contributions are sizable when the DM-particle mass is much larger
than those of weak bosons. (See also
Ref.~\cite{Hisano:2004pv}.) Further, it has been found that the SI cross
section tends to be suppressed with the $125~{\rm GeV}$ Higgs boson
mass due to an accidental cancellation. (See Fig.~5 of
Ref.~\cite{Hisano:2011cs}.) These observations indicate that both the
tree-level and the loop-level contributions are significant in a wide
range of parameter space. Taking the above discussion into account, we
calculate the scattering cross section of the neutralino with nucleon
including all the possibly dominant contributions.

For later discussion, we refer to each term in Eq.~(\ref{eff_la}) as,
\begin{eqnarray}
\frac{f_N}{m_N} &=& 
 f_{Tq}^{(N)}f_q^{\rm EWIMP}
+
\frac{3}{4} \left(q_N(2)+\bar{q}_N(2)\right)\left(g_q^{(1)}+g_q^{(2)}\right)
-\frac{8\pi}{9\alpha_s}f_{TG}^{(N)} f_G^{\rm EWIMP} 
\nonumber \\ &&
+\left(f_{Tq}^{(N)} f_q^H -\frac{8\pi}{9\alpha_s}f_{TG}^{(N)} f_G^H \right)
\nonumber \\
&\equiv& \left[
{\rm (scalar)} +  {\rm (twist\mathchar`-2)} + {\rm (gluon)} + {\rm (Higgs)}
\right]/m_N.
\label{def_fN}
\end{eqnarray}
Here ``scalar'', ``twist-2'' and ``gluon'' contributions are from
diagrams in Fig.~\ref{fig:Loop}, while we define ``Higgs''
contribution, which contains both $f^H_q$ and $f^H_G$, as shown in
Fig.~\ref{fig:Higgs}.

Before turning to numerical calculations, we briefly discuss the
gaugino-sfermion-fermion couplings, which we call the gaugino couplings
hereafter, in the high-scale SUSY scenario. The gaugino couplings are
equal to the gauge couplings at the energy scale larger than $M_{\rm SUSY}$.
With the scalar
particles decoupled at $M_{\rm SUSY}$, however, the
effective theory below the scale is not supersymmetric anymore; thus the
gaugino couplings might in general deviate from the relations. 
This deviation affects the neutralino mass matrix, leading to
corrections to $Z_{ij}$. Using the renormalization group
equations for the gaugino couplings in the split SUSY scenario given
in Ref.~\cite{Giudice:2004tc}, we explicitly calculate the
running of the couplings and find that the deviation of gaugino couplings
from the corresponding gauge couplings is less than a few percent; {\it
e.g.}, the U(1)$_Y$ gaugino coupling decreases from the supersymmetric
one by around $7\%$, while the SU(2)$_L$ gaugino coupling increases by
about $1\%$, when $M_{\rm SUSY}$ is $10^3~{\rm TeV}$ and $\tan
\beta=1$, which gives the Higgs mass of around 125~GeV.

\begin{figure}[t]
  \begin{center}
    \includegraphics[scale=1]{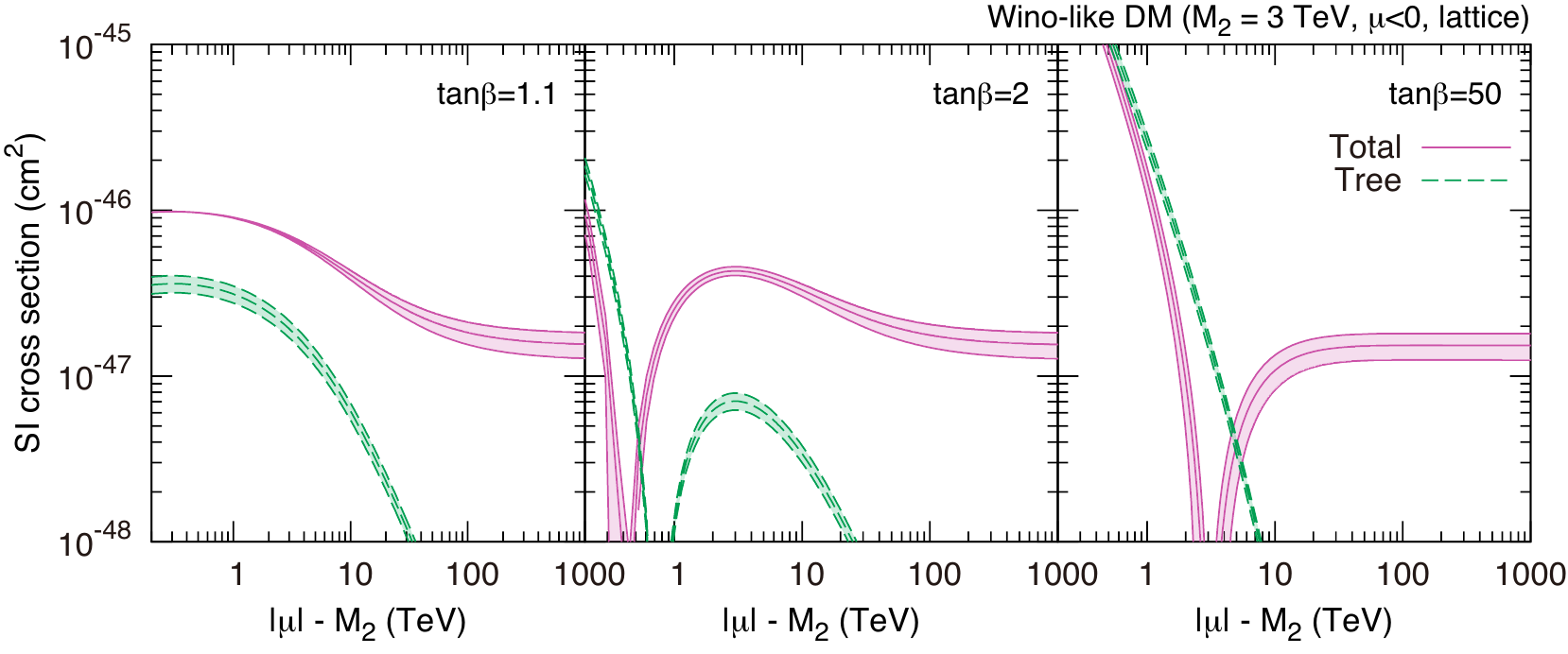}
    \includegraphics[scale=1]{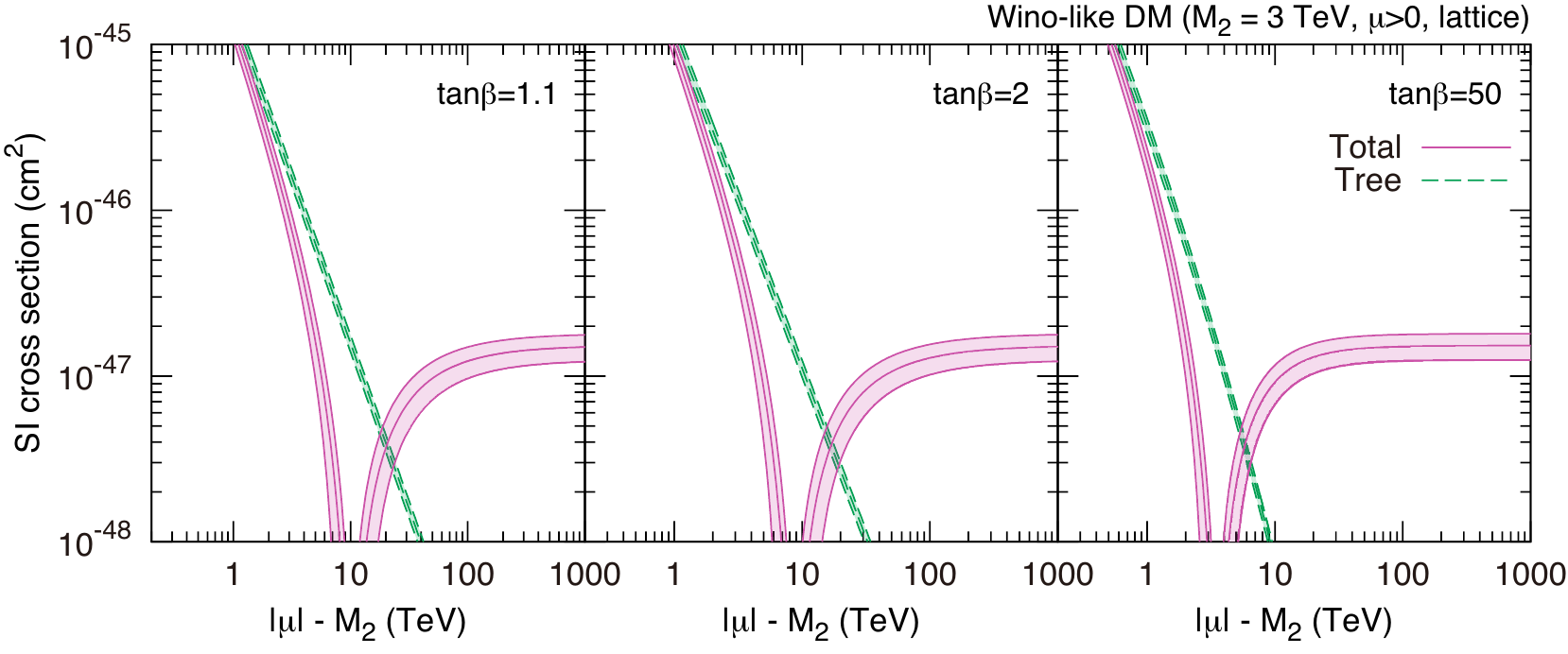}
  \end{center}
  \caption{SI cross section of Wino-like neutralino with proton.  We
    take $M_2=3~{\rm TeV}$, $m_h=125~{\rm GeV}$ and $\mu<0$ (top) and
    $\mu>0$ (bottom).  $\tan \beta$ is taken as 1.1 (left), 2 (middle)
    and 50 (right) in each panel.  Purple solid lines show the result
    from full calculation, while the result only using the Higgs
    contribution is in green dashed lines (color online). Shaded
      regions show error from the mass fractions of light quarks
      evaluated with the lattice QCD simulation.}
  \label{fig:W3000SIlatt}
\end{figure}

\begin{figure}[t]
  \begin{center}
    \includegraphics[scale=1]{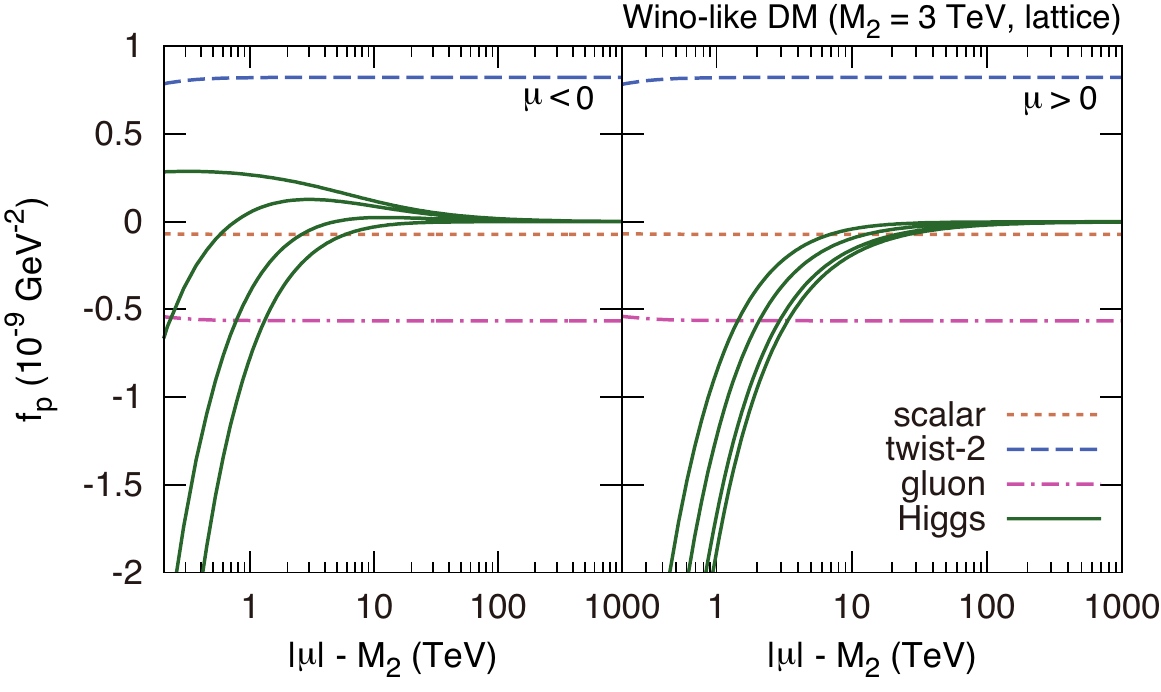}
  \end{center}
  \caption{Each contribution in the effective coupling for Wino-like
    neutralino. $\mu<0$ (left) and $\mu>0$ (right) and the other parameters
    are the same as those in Fig.~\ref{fig:W3000SIlatt}. In the panel
    ``scalar'' (orange dotted), ``twist-2'' (blue dashed), ``gluon''
    (purple dot-dashed) and ``Higgs'' (green solid) contributions
    defined in Eq.~(\ref{def_fN}) are given. For scalar, twist-2 and
    gluon contributions we take $\tan \beta=50$, while Higgs
    contribution is given for $\tan \beta=1.1$, $2$, $5$ and 50 from
    top to bottom (from bottom to top) in the left (right) panel.}  
  \label{fig:W3000fplatt}
\end{figure}


Now we are ready to give numerical results of the scattering cross
section. Figure~\ref{fig:W3000SIlatt} shows the results of the SI
scattering for the Wino-like neutralino. Here we give the cross
section of the neutralino with proton.  In the plots we take
$M_2=3~{\rm TeV}$ and $m_h=125~{\rm GeV}$ and $\mu<0$ ($\mu>0$) for the
top (bottom) panel.  $\tan \beta = 1.1$, $2$ and $50$ are taken in
left, middle and right panels, respectively.\footnote
{ Note that when $\tan \beta =1$ the tree-level axial coupling, as well as
  the tree-level Higgs coupling for the Higgsino-like DM for negative $\mu$,
  vanishes exactly. However, it is not the realistic case. In fact
  the gaugino coupling in the neutralino mass matrix receives a correction
  from renormalization-group effects from high scale, which we
  discussed above, and as a consequence the tree-level  couplings do
  not vanish.  For the purpose of studying tree- and loop-level
  contributions in general, we simply avoid $\tan \beta =1$.}
In the plot of the SI cross section, green dashed lines indicate the
SI cross section with only the Higgs contribution taken into
consideration, while purple solid lines show the result with all the
leading contributions included. Shaded regions imply error
  coming from the mass fractions.  It is found that the loop
contribution is important in a wide range of parameter space. Let us look
at the
$\mu<0$ case first.  When $\tan \beta \lesssim 2$, the Higgs
contribution scales as $\propto s^h \simeq m_W/(M_2+|\mu|)$ from
Eq.~(\ref{eq:Wino_sh}), which does not grow as $|\mu|-M_2$ gets
smaller. As a consequence, the loop contribution is comparable or
larger than the Higgs contribution, depending on $|\mu|-M_2$.  This
behavior is seen in the plot of $f_p$ in
Fig.~\ref{fig:W3000fplatt}. Here we give the plot of each
contribution defined in Eq.~(\ref{def_fN}). In the plot, the result for
$\tan \beta=50$ is given for scalar, twist-2 and gluon contributions
(though they are insensitive to $\tan\beta$), and results for
$\tan\beta=1.1$, $2$, $5$ and $50$ are given for the Higgs
contribution from top to bottom.  Since the Higgs contribution is
always positive for $\tan \beta \simeq 1.1$, it interferes
constructively with the other contributions. When $\tan \beta \gg 1$, on
the other hand, the situation gets changed.  In this case the Higgs
contribution is sensitive to $M_2$ and $|\mu|-M_2$ for given $\tan
\beta$. This fact can be seen from Eq.~(\ref{eq:Wino_sh}). Now $s^h$
is given as $s^h\simeq m_W \frac{M_2-2|\mu|/\tan
  \beta}{M_2^2-\mu^2}$. Thus the Higgs contribution is negative when
$|\mu|\lesssim M_2 \tan \beta /2$ and flips its sign in a larger $|\mu|$
region.  This causes a cancellation; as $|\mu|-M_2$ gets larger, the
absolute value of the tree-level Higgs exchanging contribution drops,
and all the negative contributions cancel the positive twist-2
contribution around $|\mu|-M_2 \sim $ a few hundred GeV to a few TeV,
depending on $\tan \beta$. We have checked that the result is almost
the same when $\tan \beta \gtrsim 5$.

When $\mu>0$, on the contrary, the Higgs contribution is always
negative. 
The right panel in Fig.~\ref{fig:W3000fplatt} shows it for
$\tan\beta=1.1$, $2$, $5$ and $50$ from bottom to top.
Thus resultant cross section is similar to those in the
case of large $\tan
\beta$ and $\mu<0$.  
In both $\mu<0$ and $\mu>0$ cases, the Higgs contribution
  becomes irrelevant and the loop contribution dominates the cross
  section when $|\mu|-M_2 \gtrsim 10~{\rm TeV}$. Then the SI cross
section lies around a value of $\sim 10^{-47}~{\rm cm}^2$, which is
consistent with the results in Refs.~\cite{Hisano:2010fy,
  Hisano:2010ct, Hisano:2011cs}.

\begin{figure}[t]
  \begin{center}
    \includegraphics[scale=1]{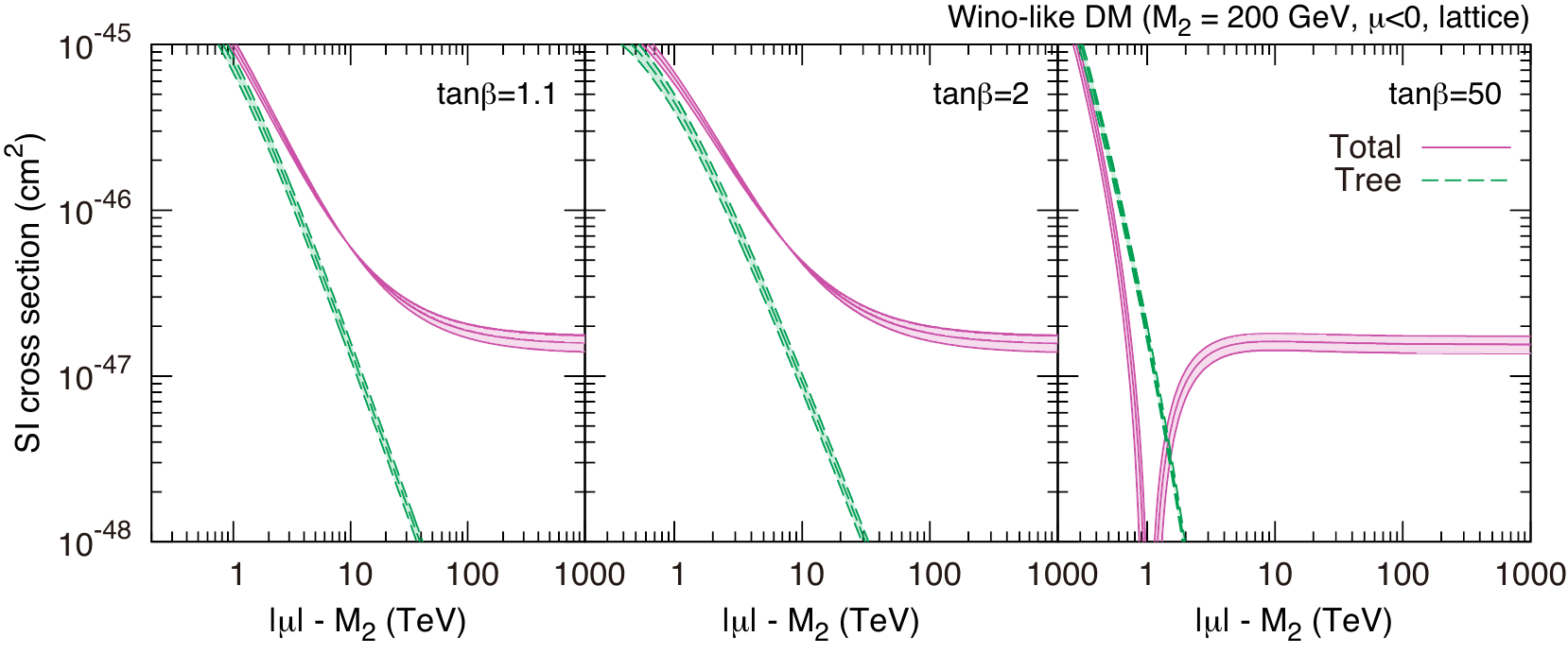}
    \includegraphics[scale=1]{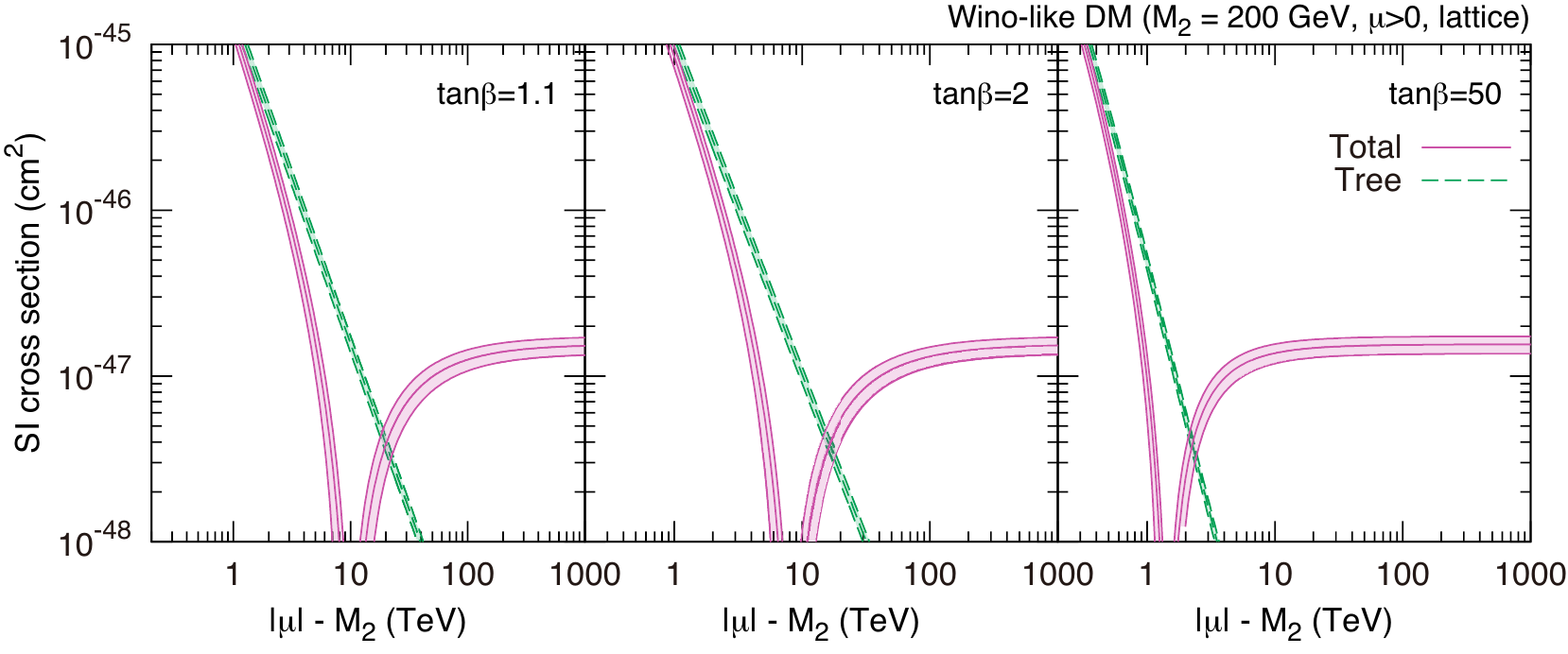}
  \end{center}
  \caption{Similar plots to those in Fig.~\ref{fig:W3000SIlatt} except
 for taking $M_2=200~{\rm GeV}$.}
  \label{fig:W0200SIlatt}
\end{figure}

\begin{figure}[t]
  \begin{center}
    \includegraphics[scale=1]{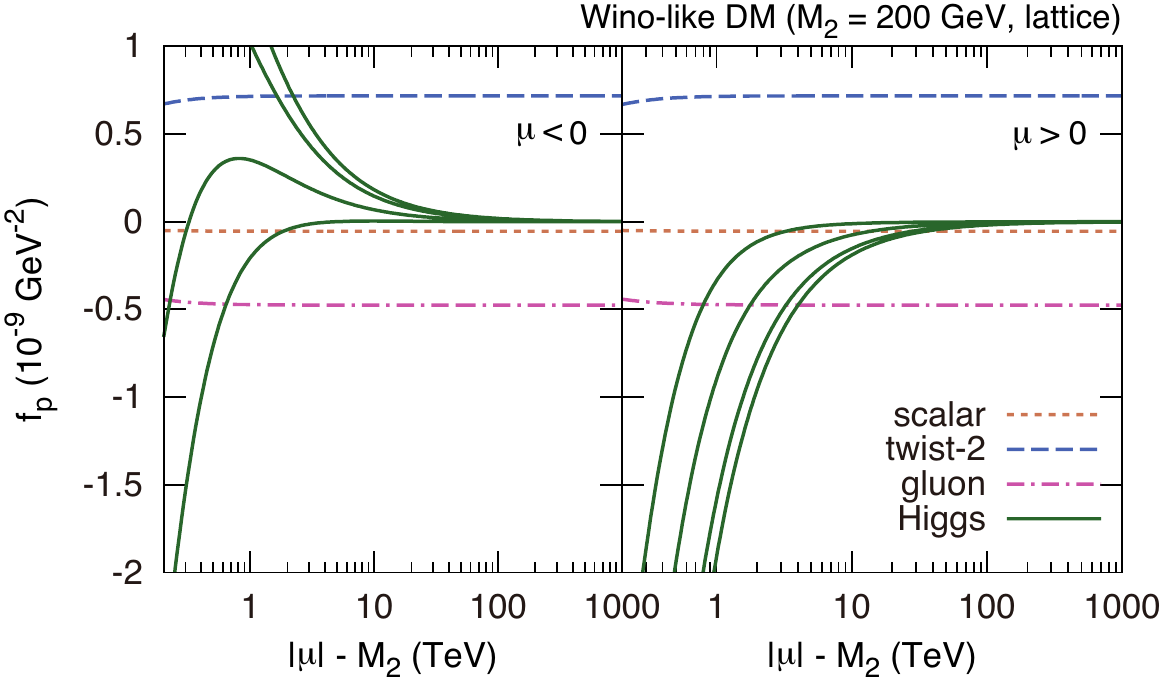}
  \end{center}
  \caption{Similar plots to those in Fig.~\ref{fig:W3000fplatt} except
 for taking $M_2=200~{\rm GeV}$.}
  \label{fig:W0200fplatt}
\end{figure}

We have checked that the SI cross section is almost independent of the
neutralino 
mass except for $\mu<0$ and low $\tan \beta$.  We also give the result
for $M_2=200~{\rm GeV}$ in Figs.~\ref{fig:W0200SIlatt} and
\ref{fig:W0200fplatt}. Here we take the other parameters the same
as those in Fig.~\ref{fig:W3000SIlatt}. In this case, the Higgs
contribution becomes larger, leading to a bit enhanced SI cross
section. However, with relatively  large $\tan \beta$, a cancellation
happens then the cross section behaves similar to the previous results,
as it is seen in the figure.

\begin{figure}[t]
  \begin{center}
    \includegraphics[scale=1]{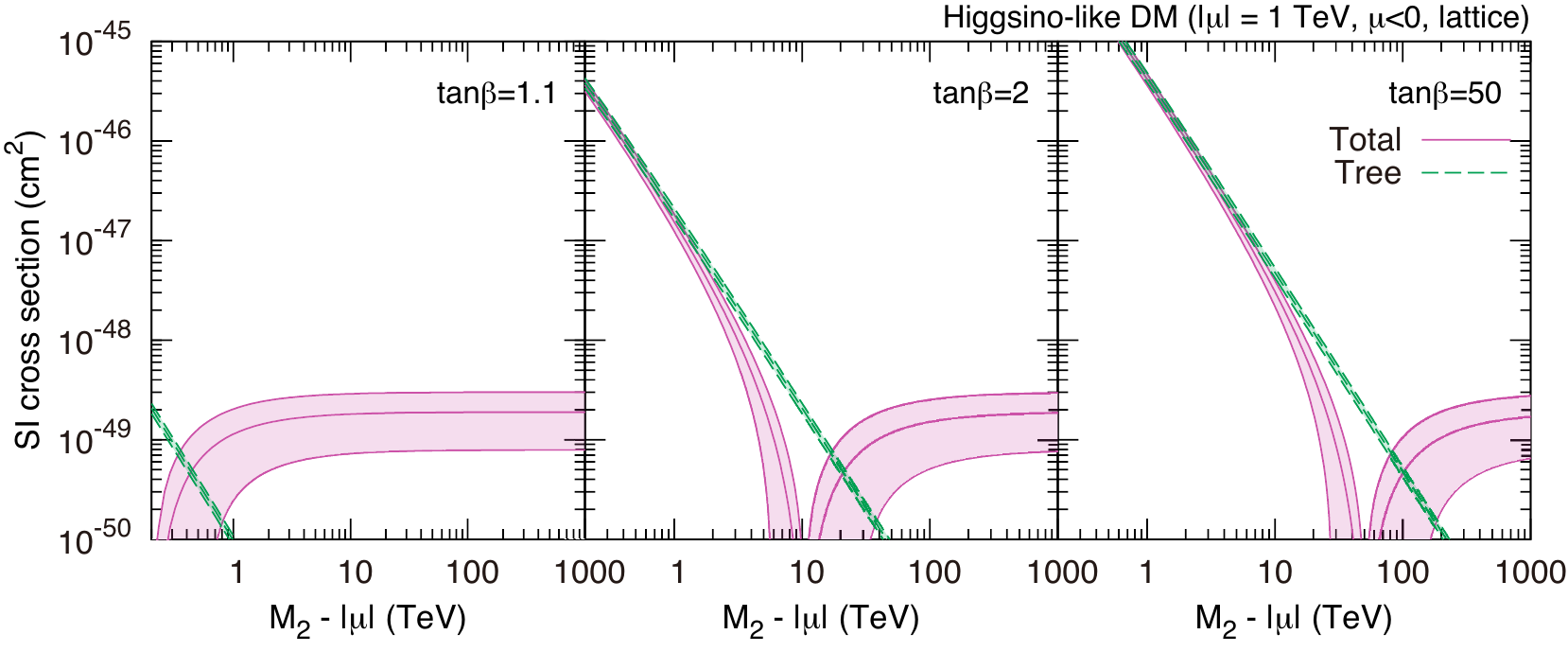}
    \includegraphics[scale=1]{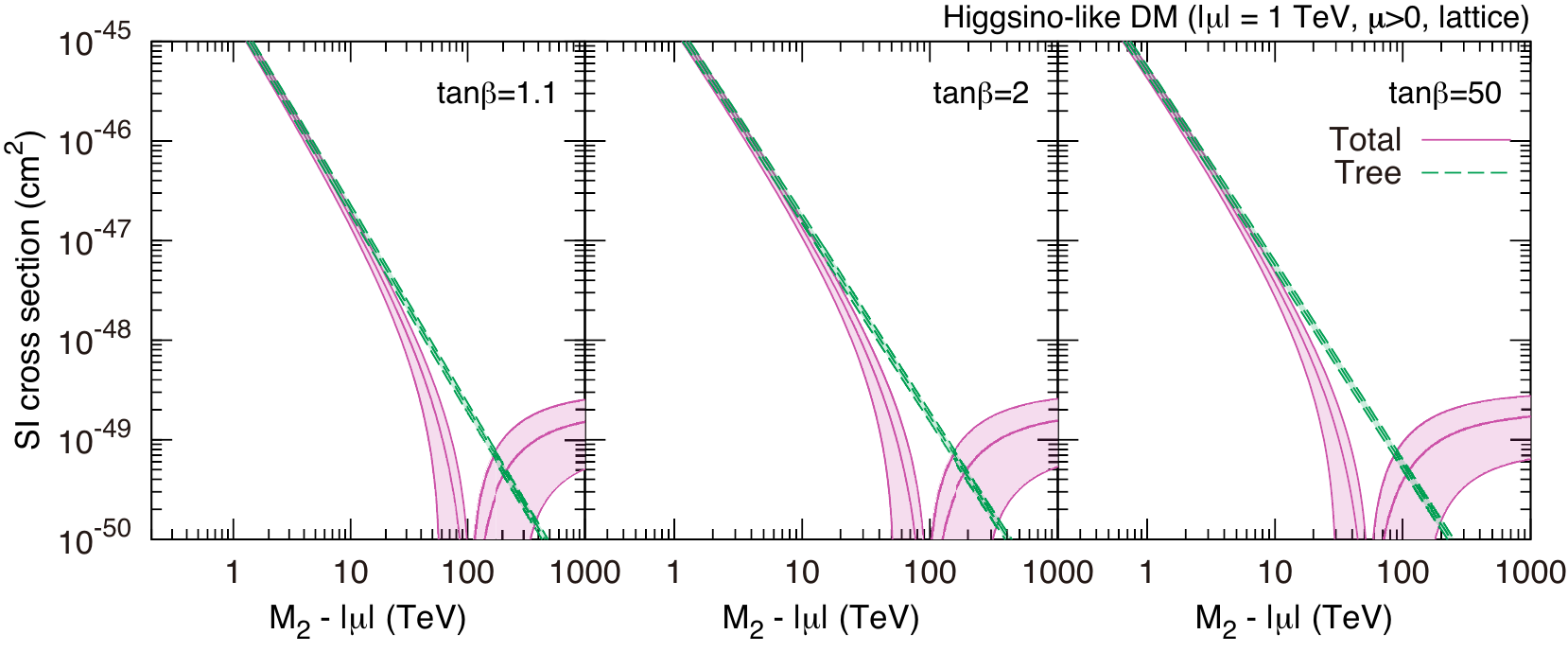}
  \end{center}
  \caption{SI cross section of Higgsino-like neutralino with proton.  We take
    $|\mu|=1~{\rm TeV}$, $m_h=125~{\rm GeV}$ and $\mu<0$ (top) and
    $\mu>0$ (bottom).  $\tan \beta$ is taken as 1.1 (left), 2 (middle)
    and 50 (right) in each panel. Line contents are the same as
    those in Fig.~\ref{fig:W3000SIlatt}.}
  \label{fig:H1000SIlatt}
\end{figure}

\begin{figure}[t]
  \begin{center}
    \includegraphics[scale=1]{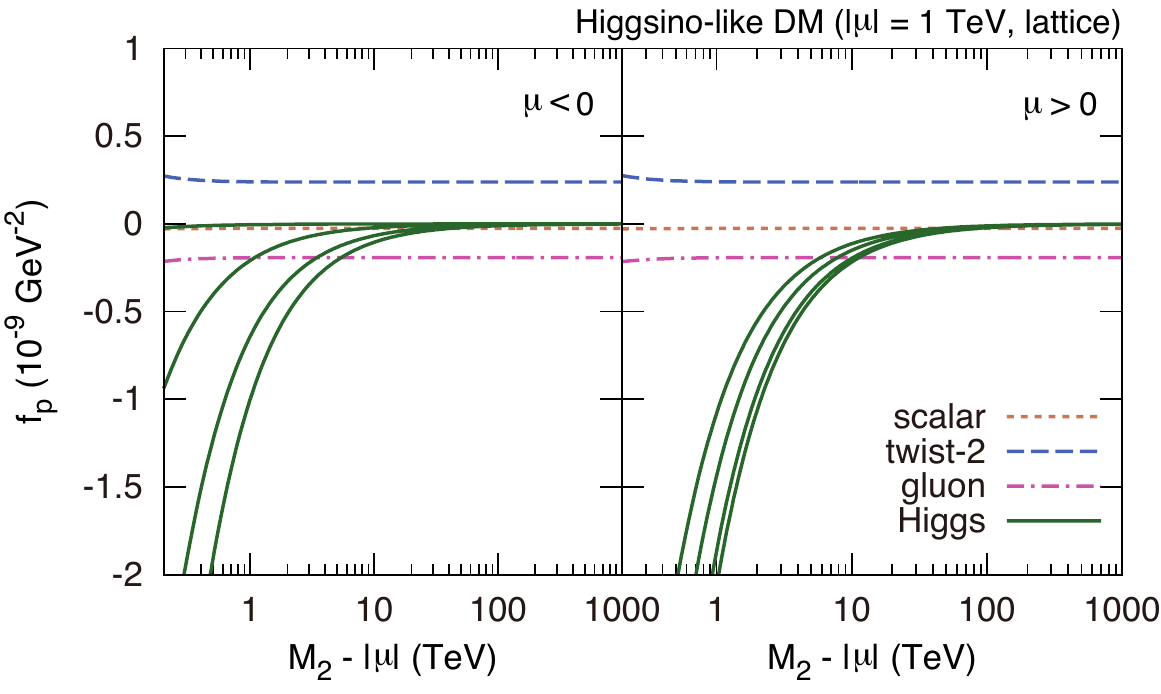}
  \end{center}
  \caption{Each contribution in the effective coupling for
    Higgsino-like neutralino. $\mu<0$ (left) and $\mu>0$ (right) and
  the other parameters as well as line contents are similar to those in
 Fig.~\ref{fig:H1000SIlatt}.}
  \label{fig:H1000fplatt}
\end{figure}

Next let us discuss the Higgsino-like neutralino case. The
results are shown in Figs.~\ref{fig:H1000SIlatt} and
\ref{fig:H1000fplatt}.  Here we take $|\mu|=1~{\rm TeV}$ and
$m_h=125~{\rm GeV}$. The upper and lower panels correspond to $\mu<0$
and $\mu>0$ cases, respectively.  $\tan \beta$ is taken as similarly
to that in Figs.~\ref{fig:W3000SIlatt} and \ref{fig:W3000fplatt}. In
the $\mu<0$ case, the Higgs contribution is suppressed by $(1-\sin 2
\beta)$ for $\tan \beta\simeq 1$, then the loop contributions become
dominant. This is clearly seen in Fig.~\ref{fig:H1000fplatt}. The
cross section is around $10^{-49}~{\rm cm}^2$ in the region $M_2 -
|\mu| \gtrsim 500~{\rm GeV}$.  When $\tan \beta$ is larger, the Higgs
contribution scales as $s^h \simeq -m_W/(M_2-|\mu|)$ and becomes
dominant in the effective coupling in the region $M_2-|\mu|\lesssim $
a few TeV to $10~{\rm TeV}$, depending on $\tan \beta$. In this case a
cancellation occurs around $M_2-|\mu|\sim $ a few dozens of TeV, and
the SI cross section is about $10^{-49}~{\rm cm}^2$ for larger values
of $M_2-|\mu|$. A similar cancellation is observed for the $\mu>0$
case. In the $\mu>0$ case, the Higgs contribution is not suppressed
around $\tan \beta \simeq 1$ in contrast to the $\mu<0$ case. That is why
a significant cancellation always happens.

Here we briefly comment on contributions by the heavy Higgs boson, which
we did not take into account. The heavy
Higgs-$\tilde{\chi}^0$-$\tilde{\chi}^0$ coupling $s^H$ is given by
$s^H\simeq -\frac{m_W\mu}{M^2_2-\mu^2} \cos 2\beta$ for the Wino-like
neutralino and $s^H\simeq \pm \frac{1}{2}
\left[\frac{m_W}{M_2-|\mu|}+\frac{m_W\tan^2
    \theta_W}{M_1-|\mu|}\right] \cos 2 \beta$ for the Higgsino-like
neutralino. Here the overall positive and negative signs correspond to the
$\mu>0$ and $\mu<0$ cases, respectively. As it is seen, contributions
from the heavy Higgs boson are suppressed when $\tan \beta \simeq 1$. Even
when $\tan \beta \gtrsim 1$, it is suppressed by the heavy Higgs mass.

\begin{figure}[t]
  \begin{center}
    \includegraphics[scale=0.8]{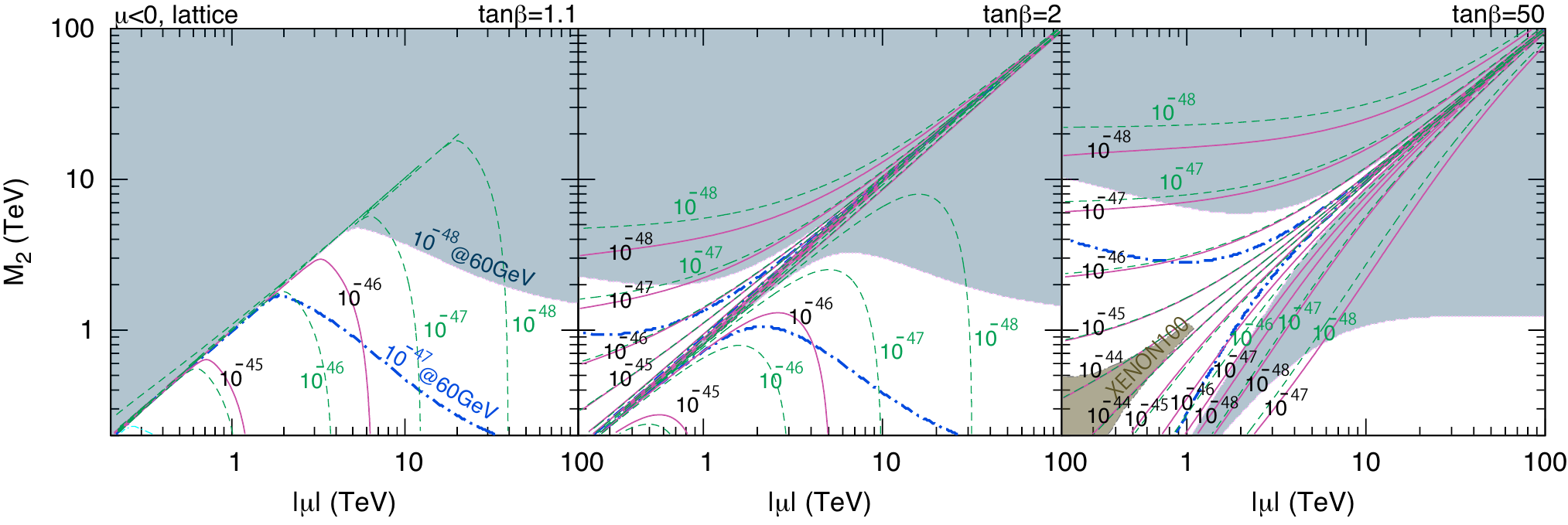}
    \includegraphics[scale=0.8]{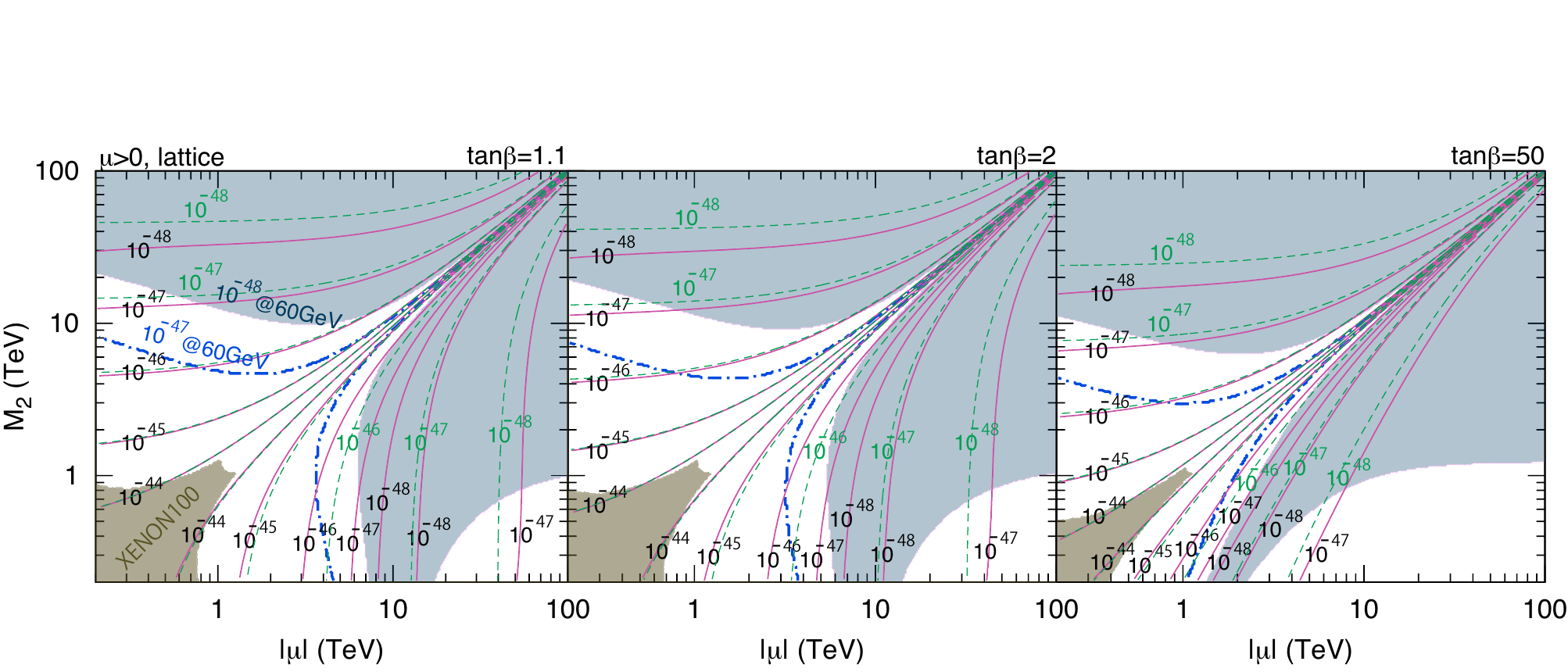}
  \end{center}
  \caption{Contour of the SI cross section in cm$^2$ unit. 
    Upper panels are in the case where $\mu<0$ and $\tan \beta=1.1$
    (left), 2 (middle), and $10$ (right) are taken, respectively. In
    the lower panels $\mu$ is set to be positive. We take
    $m_h=125~{\rm GeV}$.
 Results from full calculation and the only Higgs contribution are
 given in purple solid and green dashed lines,
 respectively. Lines are shown for the cross section larger than
 $10^{-48}~{\rm cm}^2$. Here we also show the exclusion regions by
 XENON100~\cite{Aprile:2012nq} in dark shade. Blue dot-dashed lines
 and light-shaded regions correspond to future prospects of
 experiments with sensitivities to cross section $10^{-47}~{\rm cm}^2$
 and smaller than $10^{-48}~{\rm cm}^2$ at a DM-particle mass of $\sim
 60~{\rm GeV}$, respectively.}
  \label{fig:conSI}
\end{figure}

Finally, we give contour plots of the SI cross section on the
$|\mu|-M_2$ plane in Fig.~\ref{fig:conSI}. In the plot we take
$m_h=125~{\rm GeV}$. Upper and lower panels are for the cases of
$\mu<0$ and $\mu>0$, respectively, and $\tan \beta$ is taken as 1.1, 2
and 50 from left to right in each panel.  Purple solid lines are
contours of the SI cross section of full calculation and green
dashed lines show the ones given by the Higgs contribution only.
The contours of the cross section smaller than $10^{-49}~{\rm
  cm}^2$ are not shown here.  In the figure, dark-shaded regions are
excluded by the XENON100 experiment \cite{Aprile:2012nq}. The blue
dot-dashed lines and light shaded regions correspond to prospected
reaches of future experiments.  To evaluate those sensitivity limits, we
use a value of $10^{-47}~{\rm cm}^2$ and $10^{-48}~{\rm cm}^2$ at a
DM-particle mass of $60~{\rm GeV}$ and rescale it with respect to the
DM-particle 
mass. The former value is based on a discovery sensitivity in a ton-year
  experiment,\footnote
{For reference, $8.6\times 10^{-48}~{\rm cm}^2$ at a DM-particle mass
  of $\sim 60~{\rm GeV}$ is the sensitivity of 90\% C.L. discovery at
  a ton-year Xenon target experiment~\cite{Gutlein:2010tq}.  }
while the latter comes from the fact that a sensitivity for the
cross section of less than $10^{-48}~{\rm cm}^2$ is difficult to be
achieved due to atmospheric neutrino background
\cite{Gutlein:2010tq}. 

In the Wino-like neutralino region, it is seen that the full calculation
deviates from the one given by the Higgs contribution significantly.
This is due to the
suppression of the Higgs contribution (especially for the $\mu<0$ and
$\tan \beta \sim 1$ case) or the cancellation in the effective
coupling. In addition, as we discussed previously, the cross section
can be enhanced due to constructive interference between the Higgs and
twist-2 contributions in the $\mu<0$, $M_2 \lesssim {\rm TeV}$ and low
$\tan \beta$ region (see left and middle plot in upper panel). Thus such a
region can be probed in a future experiment even when $|\mu|$ is as
large as dozens of TeV. If much better sensitivity was
accomplished, larger $|\mu|$ could be also studied.

For the Higgsino-like case, the Higgs contribution almost determines
the cross section in the region where future experiments may reach.
The loop effect becomes important when the Higgs contribution is
suppressed (for $\mu<0$ and $\tan \beta \simeq 1$ case) or in the
region where $M_2$ is above several TeV. For the $\mu<0$ and $\tan
\beta \simeq 1$ case, the cross section is around $10^{-49}~{\rm
  cm}^2$ thus it is far below the sensitivity of future experiments.

\section{Conclusion}
\label{sec:conclusion}

The high-scale supersymmetry, in which SUSY particles are much heavier
than the weak scale except for gauginos and/or Higgsinos, is favored
from viewpoints of the discovered 125~GeV Higgs boson, null results in
the SUSY 
particle searches at the LHC, and the SUSY FCNC and CP problems.  In this
scenario, the Wino-like or Higgsino-like neutralino is a good candidate
for the dark matter in the Universe. While a Wino with a mass of
2.7--3.0~TeV or a Higgsino with a mass of 1~TeV is predicted in the
thermal relic scenario, the non-thermal production may explain the
observed dark matter abundance in even lighter mass. In this
scenario, the elastic scattering of the neutralino with nucleon,
relevant to the direct dark matter search experiments, is induced by
tree-level Higgs boson exchange diagrams and also loop diagrams due to
the electroweak interaction.

In this article, we evaluate the SI cross section of the Wino-like or
Higgsino-like neutralino including contribution from the loop diagrams
of the weak bosons. Since the loop diagrams are not suppressed by the
neutralino mass, they may be comparable to or even dominate over the
Higgs-exchange contribution, especially for the Wino-like
neutralino. As a result, the SI cross section is sensitive to the sign
of $\mu$ and $\tan\beta$, in addition to absolute values of the
Higgsino and Wino mass parameters since the diagrams
constructively or destructively interfere with each other.

Because of atmospheric neutrino background, it is difficult to
  discover the DM in the direct detection experiments when the SI cross
  section is smaller than $10^{-48}~{\rm cm}^2$ at a DM-particle mass
  around $60~{\rm GeV}$.
We found that the prediction for the SI cross section is larger than
the limit in a broad parameter region. (See Fig.~\ref{fig:conSI}.) 
The large-scale experiments for the direct DM detection are hopeful.

\section*{Acknowledgments}

We would like to thank Tomohiro Takesako for collaboration and
discussion in the early stages of this work. We are also grateful to
Satoshi Shirai for useful comments.  This work is supported by
Grant-in-Aid for Scientific research from the Ministry of Education,
Science, Sports, and Culture (MEXT), Japan, No. 20244037,
No. 20540252, No. 22244021 and No.23104011 (JH), and also by World
Premier International Research Center Initiative (WPI Initiative),
MEXT, Japan. This work was also supported in part by the
U.S. Department of Energy under contract No. DE-FG02-92ER40701, and by
the Gordon and Betty Moore Foundation (KI).  The work of NN is
supported by Research Fellowships of the Japan Society for the
Promotion of Science for Young Scientists.

\appendix

\section{ Spin-independent Cross Section Evaluated with Mass Fraction
  from Chiral Perturbation}

\begin{figure}[t]
  \begin{center}
    \includegraphics[scale=1]{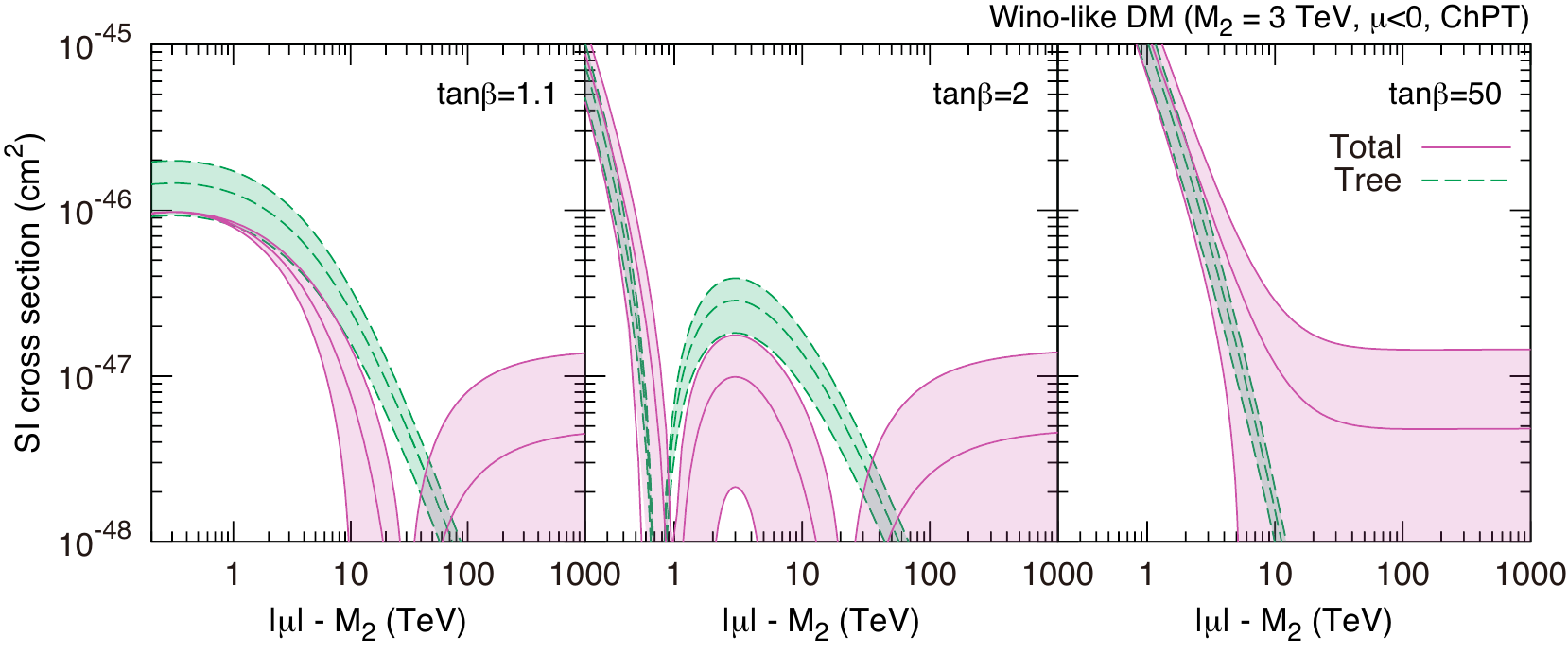}
    \includegraphics[scale=1]{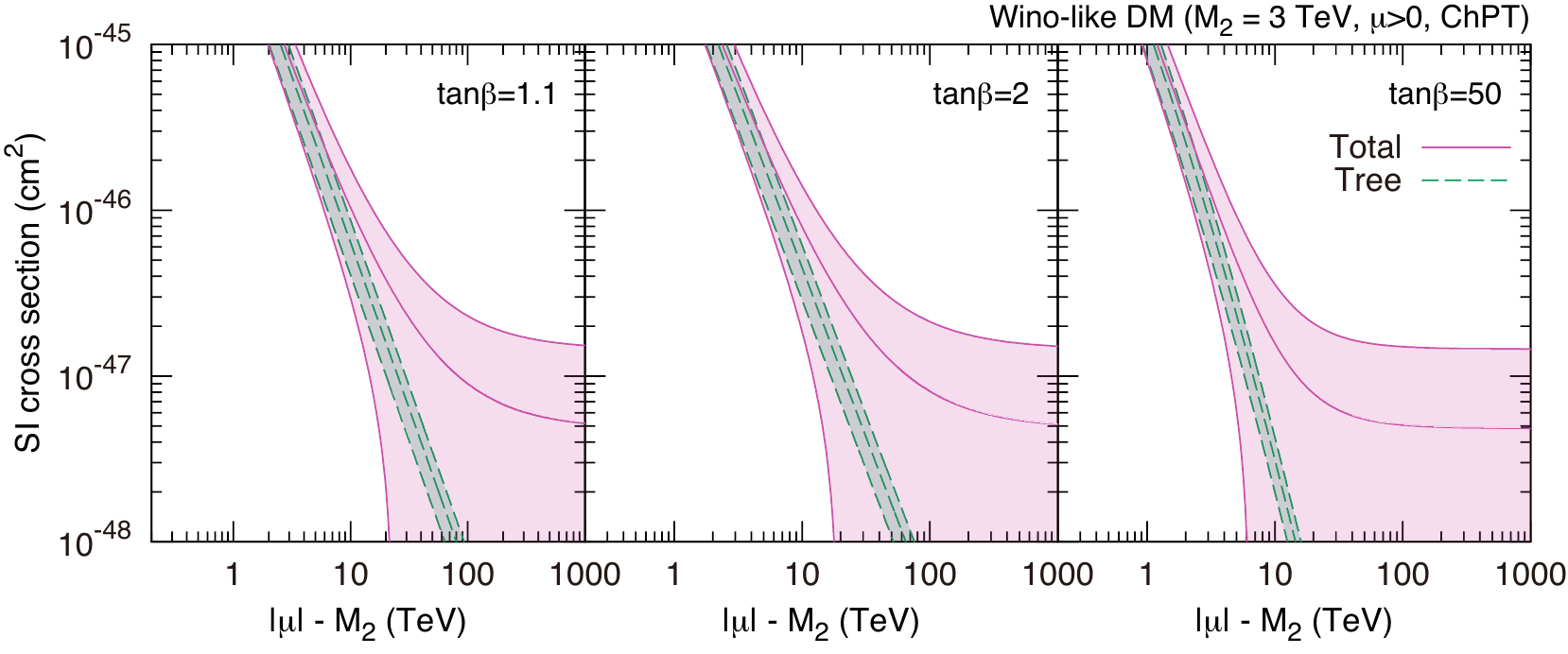}
  \end{center}
  \caption{Similar plots to those in Fig.~\ref{fig:W3000SIlatt} except that 
the mass fractions $f_{Tq}^{(N)}$ from the ChPT are used.}
  \label{fig:W3000SIchpt}
\end{figure}

\begin{figure}[t]
  \begin{center}
    \includegraphics[scale=1]{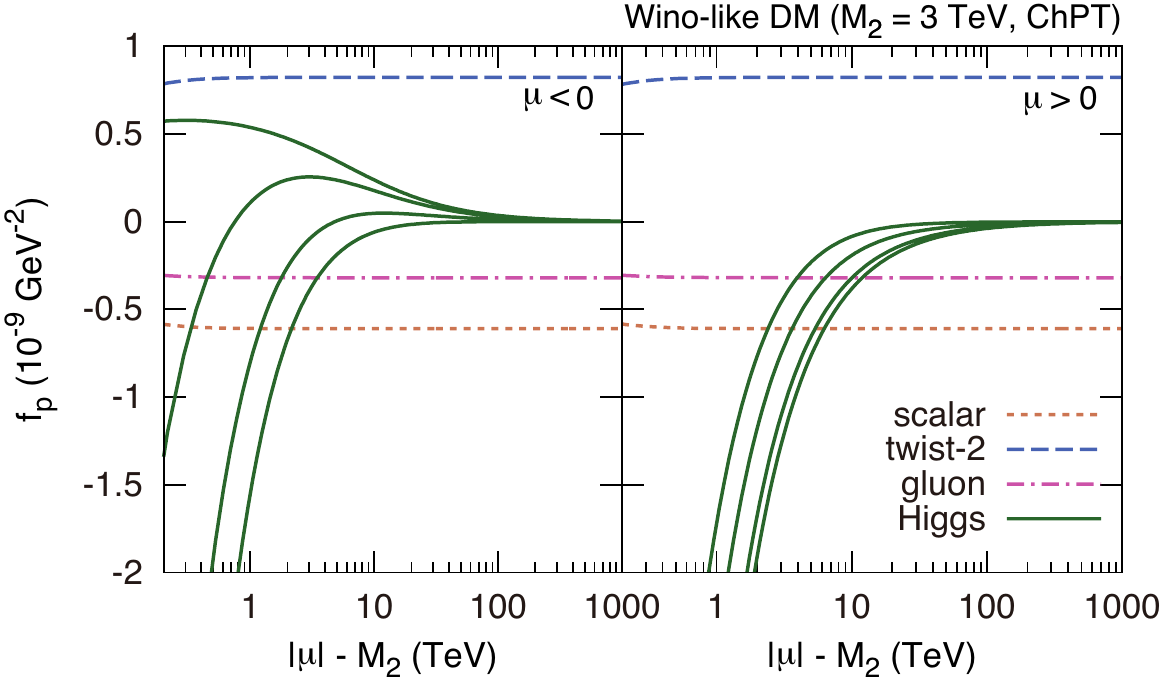}
  \end{center}
  \caption{Similar plots to those in Fig.~\ref{fig:W3000fplatt} except
 that the mass fractions $f_{Tq}^{(N)}$ from the ChPT are used.}
  \label{fig:W3000fpchpt}
\end{figure}

\begin{figure}[t]
  \begin{center}
    \includegraphics[scale=1]{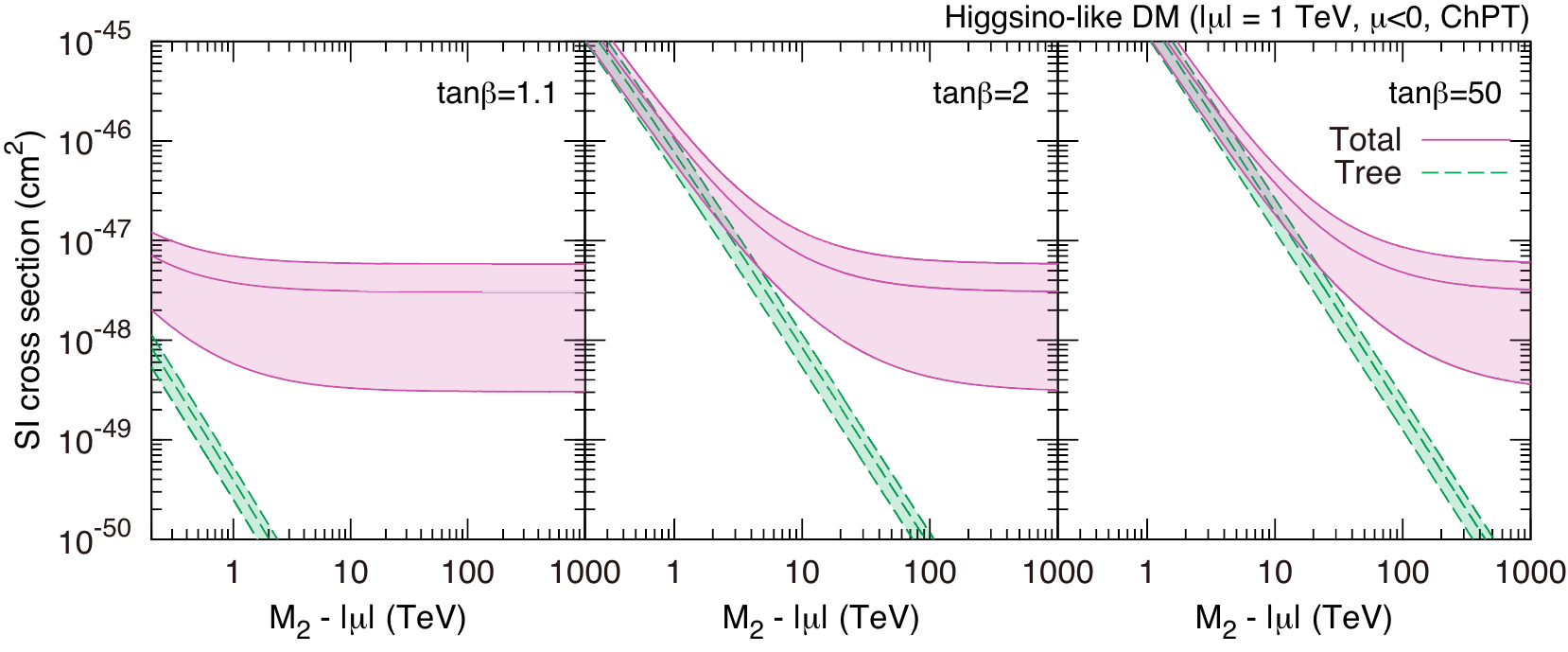}
    \includegraphics[scale=1]{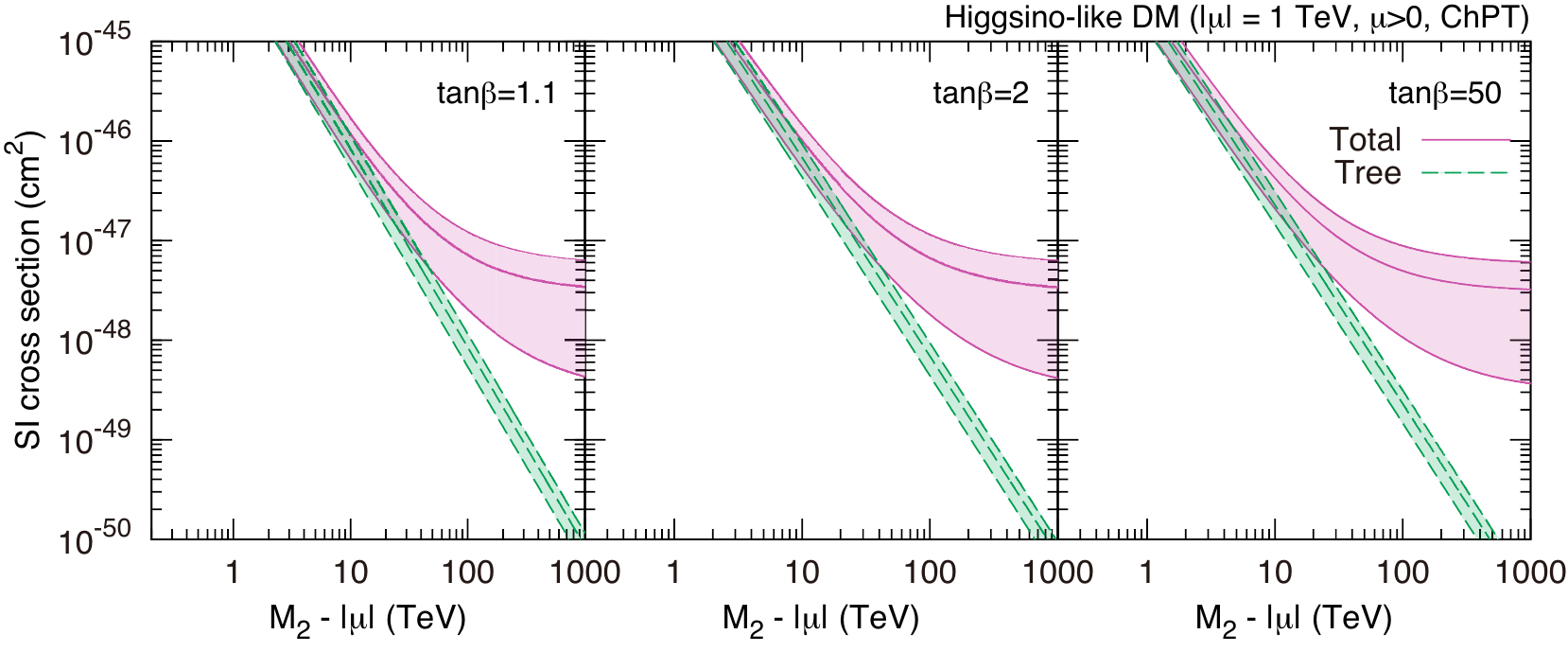}

  \end{center}
  \caption{Similar plots to those in Fig.~\ref{fig:H1000SIlatt} except
 that the mass fractions $f_{Tq}^{(N)}$ from the ChPT are used.}
  \label{fig:H1000SIchpt}
\end{figure}

\begin{figure}[t]
  \begin{center}
    \includegraphics[scale=1]{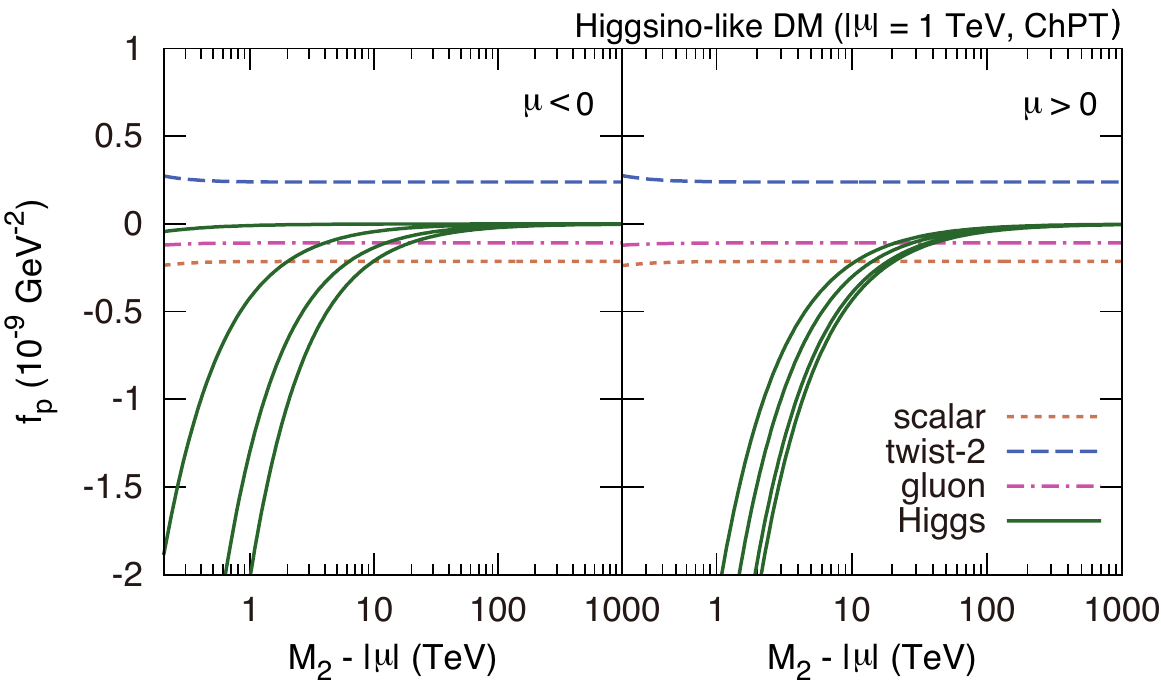}
  \end{center}
  \caption{Similar plots to those in Fig.~\ref{fig:H1000fplatt} except
 that the mass fractions $f_{Ts}^{(N)}$ from the ChPT are used.}
  \label{fig:H1000fpchpt}
\end{figure}

In this paper we have used the input parameters extracted from the lattice
QCD simulations for the mass fractions $f_{Tq}^{(N)}$. As a result,
the error of the calculation is small, as we have seen above.
However, another result is also reported for the mass fractions based
on the chiral perturbation theory. In this case the mass fractions for
proton are given as $f^{(p)}_{Tu}=0.024(4)$, $f^{(p)}_{Td}=0.041(6)$
and $f^{(p)}_{Ts}=0.40(14)$~\cite{Pavan:2001wz, Borasoy:1996bx}. A large
discrepancy\footnote{
However, a recent calculation based on the covariant baryon chiral
perturbation theory in Ref.~\cite{Alarcon:2012nr} gives a smaller value
for the strangeness content of nucleon than those in the previous works.
Indeed, it is consistent with the lattice results, while its error
is much larger than those with the lattice simulations.
} is seen for $f^{(p)}_{Ts}$, as well as larger error.  To
see the impact of the mass fractions on the SI cross section, we plot
the results using the mass fractions extracted from the ChPT (and the
other parameters are unchanged) in Figs.~\ref{fig:W3000SIchpt} and
\ref{fig:W3000fpchpt} for the Wino-like case, and
Figs.~\ref{fig:H1000SIchpt} and \ref{fig:H1000fpchpt} for the
Higgsino-like case. In both cases we find that the theoretical error
of the cross section is much larger than those presented in
Figs.~\ref{fig:W3000SIlatt} and \ref{fig:H1000SIlatt}.  Let us look at the
Wino-like neutralino case, for example.  The SI cross section has
error of an order of magnitude when $|\mu| - M_2\gtrsim 10~{\rm TeV}$ (a
few TeV) for $\tan \beta=1.1$ (50), and what is worse, the lower value
is undetermined for the larger values of $|\mu|$.  In the
Higgsino-like case, it is seen that the error is much larger than those in
the result which is based on the lattice QCD simulation. Therefore, we
conclude that in using the input of the mass fractions based on the
ChPT, the SI cross section cannot be predicted due to the large
uncertainty.

\section{Spin-dependent  Cross Section}

For completeness, we show the results for the SD cross section. 
In the case of the SD scattering, the tree-level
axial vector coupling is induced through the $Z$ boson
exchange. With the loop-level contribution combined, the axial vector
coupling is given as
\begin{eqnarray}
d_q = d_q^{\rm tree} + d_q^{\rm EWIMP}  . 
\end{eqnarray}
Here $d_q^{\rm EWIMP}$ is taken from Eq.~(4.3) in
Ref.~\cite{Hisano:2011cs}, and the tree-level contribution is
\begin{eqnarray}
 d_q^{\rm tree} = \frac{g_2^2}{8m_W^2}(|Z_{13}|^2-|Z_{14}|^2)~T^3_q,
\label{eq:dqtree}
\end{eqnarray}
with $T^3_q$ the weak isospin of light quarks.  Then as in $f_N$, we call 
each term in Eq.~(\ref{an}), 
\begin{eqnarray}
a_N &=& d^{\rm tree}\Delta q_N  + d^{\rm EWIMP}\Delta q_N 
\nonumber \\
&\equiv&  {\rm (axial\mathchar`-tree)} + {\rm (axial\mathchar`-loop)}.
\label{def_aN}
\end{eqnarray}
The first term is derived from the $W/Z$ box diagrams shown in
Fig.~\ref{fig:Loop}, while the second term is given in the tree-level $Z$
exchange (the right diagram in Fig.~\ref{fig:Higgs}).

\begin{figure}[t]
  \begin{center}
    \includegraphics[scale=1]{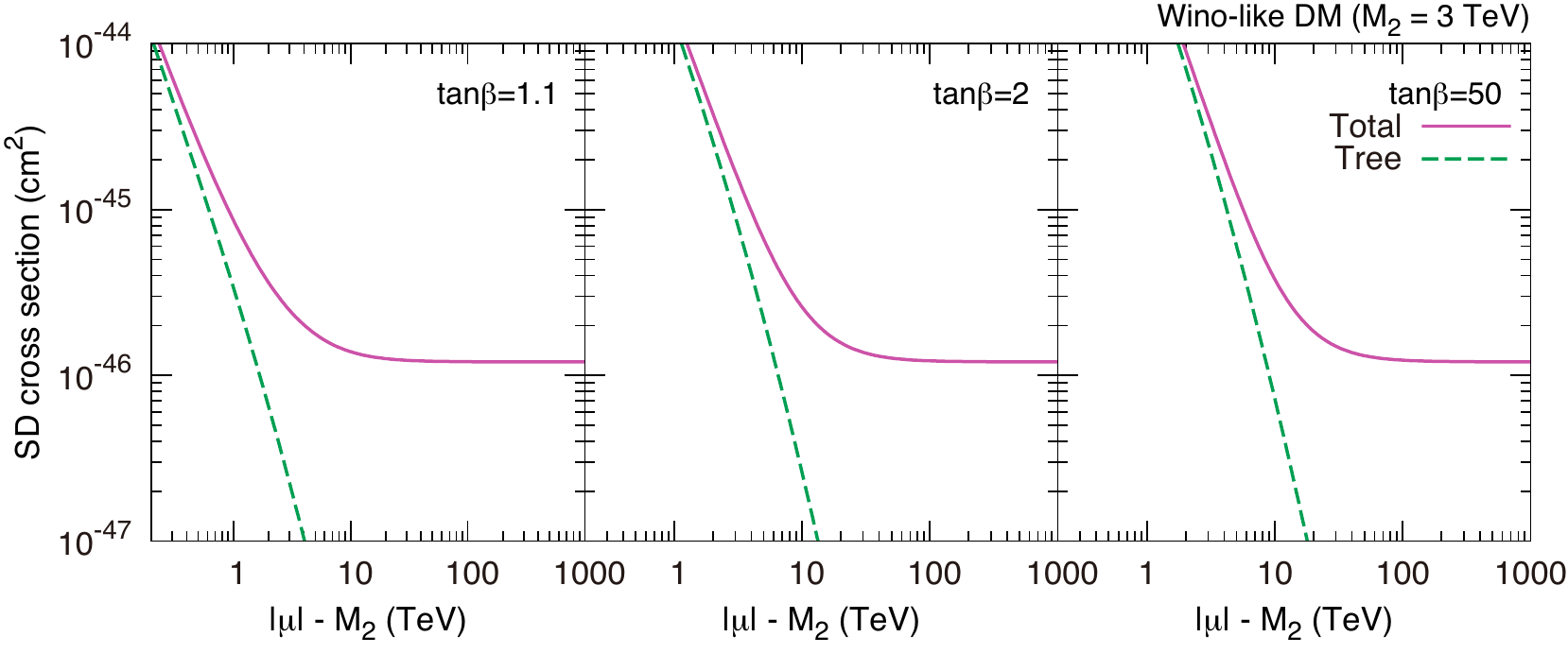}
    \includegraphics[scale=1]{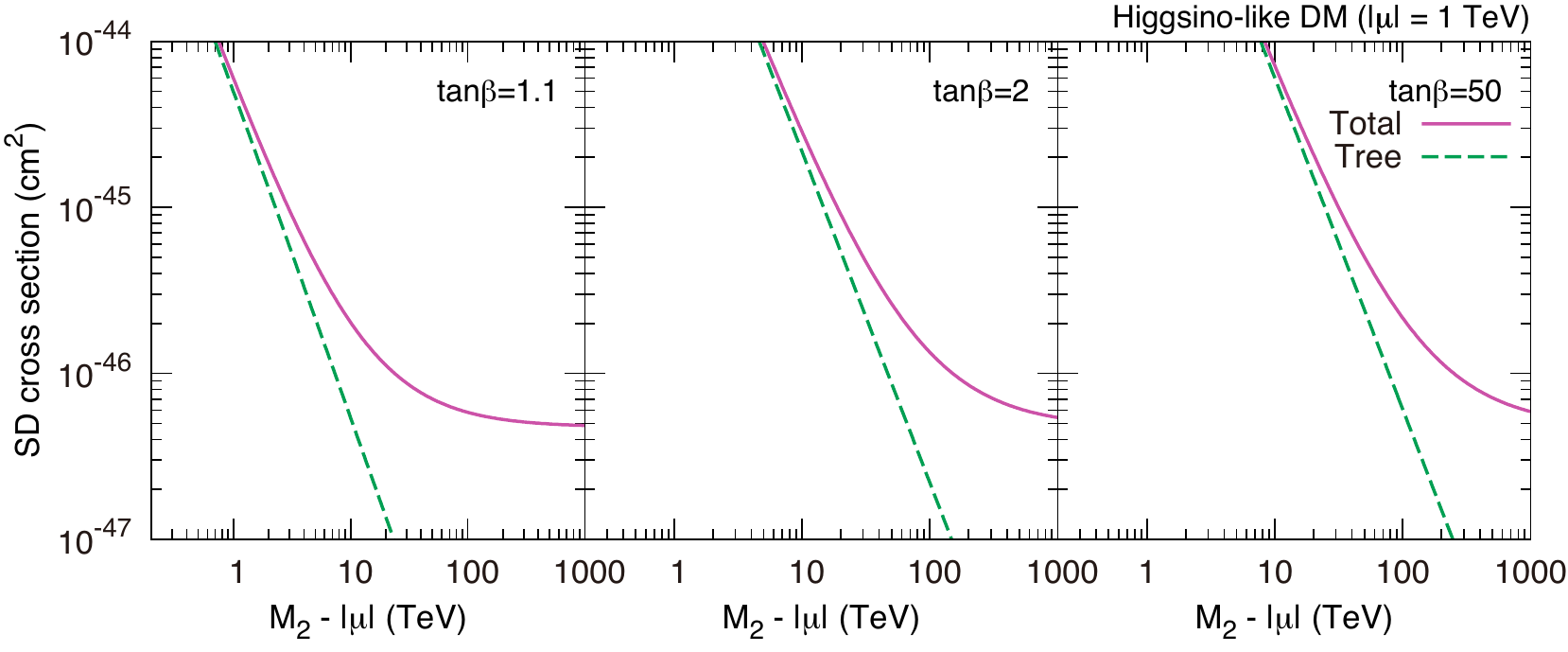}
  \end{center}
  \caption{SD cross section for Wino-like neutralino (top) and
    Higgsino-like neutralino (bottom) where the input parameters are
    set to be the same as those in Figs.~\ref{fig:W3000SIlatt} and
    \ref{fig:H1000SIlatt}, respectively. Results are shown for
    one-loop level and tree-level in purple solid and green dashed
    lines, respectively. }
  \label{fig:SD}
\end{figure}

\begin{figure}[t]
  \begin{center}
    \includegraphics[scale=1]{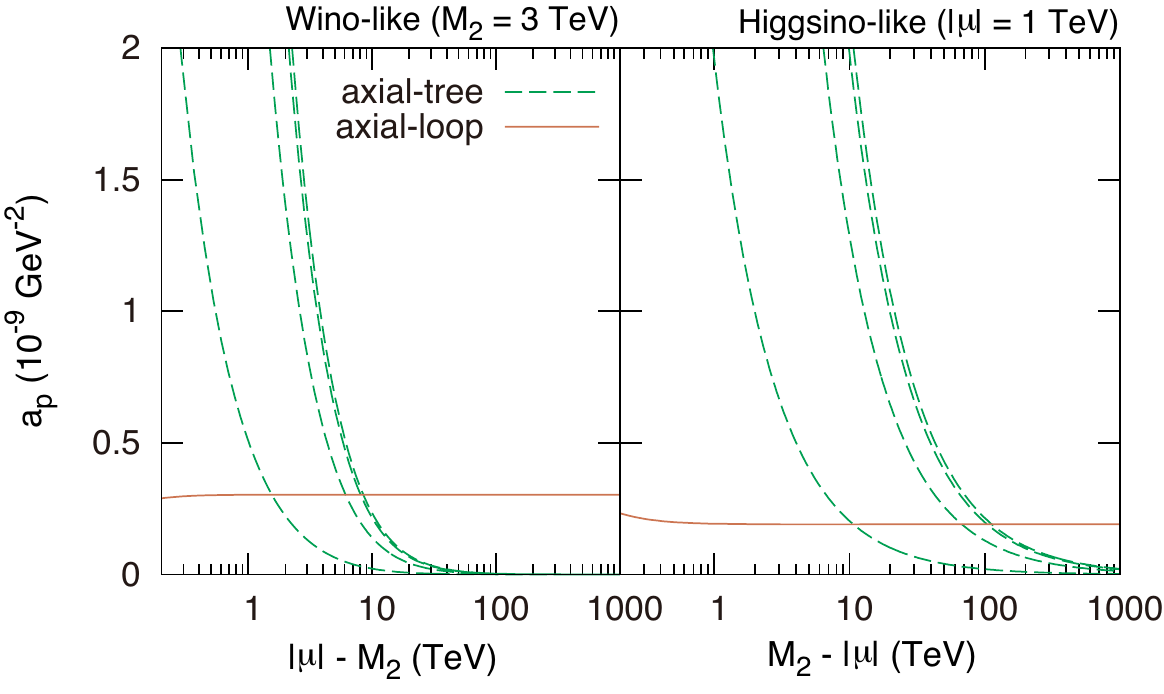}
  \end{center}
  \caption{Each contribution in the effective axial coupling for
    Wino-like (left) and Higgsino-like neutralinos (right).  In the
    figure tree-level and one-loop level contributions, defined in
    Eq.~(\ref{def_aN}), are shown in green dashed and orange solid
    lines, respectively. For tree-level contribution $\tan \beta =1.1$
    2, 5 and 50 are taken from bottom to top.}
  \label{fig:ap}
\end{figure}

The SD cross section for the Wino-like and
Higgsino-like cases are presented in Fig.~\ref{fig:SD}, while
the effective axial coupling is given in Fig.~\ref{fig:ap}. In the plots
we take the same values for the SUSY parameters as those in
Figs.~\ref{fig:W3000SIlatt} and \ref{fig:H1000SIlatt} for the Wino-like
and Higgsino-like cases, respectively. As is obvious from
Eq.~(\ref{eq:dqtree}), the result is independent of the sign of $\mu$.
In both cases, tree-level and loop-level couplings are constructive.
While the tree-level contribution highly depends on $\tan \beta$ when
$\tan \beta \lesssim 10$, it turns out to be insensitive to $\tan
\beta$ otherwise. 
The SD cross section obtained is so small that there is little
hope to detect DM via the SD interactions in future experiments.

{}


\begin{thebibliography}{99}

\bibitem{:2012rz} 
  G.~Aad {\it et al.}  [ATLAS Collaboration],
  arXiv:1208.0949 [hep-ex];\\
  S.~Chatrchyan {\it et al.}  [CMS Collaboration],
  arXiv:1207.1898 [hep-ex].

\bibitem{:2012gk} 
  G.~Aad {\it et al.}  [ATLAS Collaboration],
  Phys.\ Lett.\ B {\bf 716}, 1 (2012); \\
%
  S.~Chatrchyan {\it et al.}  [CMS Collaboration],
  Phys.\ Lett.\ B {\bf 716}, 30 (2012)
.

\bibitem{Okada:1990vk} 
  Y.~Okada, M.~Yamaguchi and T.~Yanagida,
  Prog.\ Theor.\ Phys.\  {\bf 85}, 1 (1991);\\
  Y.~Okada, M.~Yamaguchi and T.~Yanagida,
  Phys.\ Lett.\ B {\bf 262}, 54 (1991);\\
  H.~E.~Haber and R.~Hempfling,
  Phys.\ Rev.\ Lett.\  {\bf 66}, 1815 (1991);\\
  J.~R.~Ellis, G.~Ridolfi and F.~Zwirner,
  Phys.\ Lett.\ B {\bf 257}, 83 (1991);\\
  J.~R.~Ellis, G.~Ridolfi and F.~Zwirner,
  Phys.\ Lett.\ B {\bf 262}, 477 (1991).

\bibitem{Hall:2011aa} 
  L.~J.~Hall, D.~Pinner and J.~T.~Ruderman,
  JHEP {\bf 1204}, 131 (2012); \\
%
  A.~Arbey, M.~Battaglia, A.~Djouadi, F.~Mahmoudi and J.~Quevillon,
  Phys.\ Lett.\ B {\bf 708}, 162 (2012); \\
  P.~Draper, P.~Meade, M.~Reece and D.~Shih,
  Phys.\ Rev.\ D {\bf 85}, 095007 (2012).


\bibitem{Wells:2003tf} 
  J.~D.~Wells,
  hep-ph/0306127.


\bibitem{ArkaniHamed:2004fb} 
  N.~Arkani-Hamed and S.~Dimopoulos,
  JHEP {\bf 0506}, 073 (2005). 

\bibitem{Giudice:2004tc} 
  G.~F.~Giudice and A.~Romanino,
  Nucl.\ Phys.\ B {\bf 699}, 65 (2004)
  [Erratum-ibid.\ B {\bf 706}, 65 (2005)].

\bibitem{ArkaniHamed:2004yi} 
  N.~Arkani-Hamed, S.~Dimopoulos, G.~F.~Giudice and A.~Romanino,
  Nucl.\ Phys.\ B {\bf 709}, 3 (2005).

\bibitem{Wells:2004di} 
  J.~D.~Wells,
  Phys.\ Rev.\ D {\bf 71}, 015013 (2005)~.


\bibitem{Hall:2009nd} 
  L.~J.~Hall and Y.~Nomura,
  JHEP {\bf 1003}, 076 (2010)
.

\bibitem{Hall:2011jd} 
  L.~J.~Hall and Y.~Nomura,
  JHEP {\bf 1201}, 082 (2012)
.

\bibitem{Giudice:2011cg} 
  G.~F.~Giudice and A.~Strumia,
  Nucl.\ Phys.\ B {\bf 858}, 63 (2012);\\
  M.~Ibe and T.~T.~Yanagida,
  Phys.\ Lett.\ B {\bf 709}, 374 (2012);\\
  M.~Ibe, S.~Matsumoto and T.~T.~Yanagida,
  Phys.\ Rev.\ D {\bf 85}, 095011 (2012).

\bibitem{Gabbiani:1996hi} 
  F.~Gabbiani, E.~Gabrielli, A.~Masiero and L.~Silvestrini,
  Nucl.\ Phys.\ B {\bf 477}, 321 (1996)
.

\bibitem{Fukugita:1986hr} 
  M.~Fukugita and T.~Yanagida,
  Phys.\ Lett.\ B {\bf 174}, 45 (1986);\\
  W.~Buchmuller, R.~D.~Peccei and T.~Yanagida,
  Ann.\ Rev.\ Nucl.\ Part.\ Sci.\  {\bf 55}, 311 (2005).

\bibitem{Hisano:2012wq} 
  J.~Hisano, D.~Kobayashi and N.~Nagata,
  Phys.\ Lett.\ B {\bf 716}, 406 (2012).



\bibitem{Jeong:2011sg} 
  K.~S.~Jeong, M.~Shimosuka and M.~Yamaguchi,
  arXiv:1112.5293 [hep-ph];\\
  R.~Saito and S.~Shirai,
  Phys.\ Lett.\ B {\bf 713}, 237 (2012);\\
  R.~Sato, S.~Shirai and K.~Tobioka;
  arXiv:1207.3608 [hep-ph];\\
  B.~Bhattacherjee, B.~Feldstein, M.~Ibe, S.~Matsumoto and T.~T.~Yanagida,
  arXiv:1207.5453 [hep-ph];\\
  M.~Bose and M.~Dine,
  arXiv:1209.2488 [hep-ph];\\
  L.~J.~Hall, Y.~Nomura and S.~Shirai,
  arXiv:1210.2395 [hep-ph].


\bibitem{Hisano:2006nn}
  J.~Hisano, S.~Matsumoto, M.~Nagai, O.~Saito and M.~Senami,
  Phys.\ Lett.\  B {\bf 646}, 34 (2007).

\bibitem{Cirelli:2007xd} 
  M.~Cirelli, A.~Strumia and M.~Tamburini,
  Nucl.\ Phys.\ B {\bf 787}, 152 (2007)
.

\bibitem{Gherghetta:1999sw} 
  T.~Gherghetta, G.~F.~Giudice and J.~D.~Wells,
  Nucl.\ Phys.\ B {\bf 559}, 27 (1999).
.

\bibitem{Moroi:1999zb} 
  T.~Moroi and L.~Randall,
  Nucl.\ Phys.\ B {\bf 570}, 455 (2000)
.

\bibitem{Randall:1998uk} 
  L.~Randall and R.~Sundrum,
  Nucl.\ Phys.\ B {\bf 557}, 79 (1999);\\
%
  G.~F.~Giudice, M.~A.~Luty, H.~Murayama and R.~Rattazzi,
  JHEP {\bf 9812}, 027 (1998)
.

\bibitem{Murakami:2000me} 
  B.~Murakami and J.~D.~Wells,
  Phys.\ Rev.\ D {\bf 64}, 015001 (2001)
.

\bibitem{Moroi:2011ab} 
  T.~Moroi and K.~Nakayama,
  Phys.\ Lett.\ B {\bf 710}, 159 (2012)
.

\bibitem{Hisano:2010fy} 
  J.~Hisano, K.~Ishiwata and N.~Nagata,
  Phys.\ Lett.\ B {\bf 690}, 311 (2010)
.

\bibitem{Hisano:2010ct}
  J.~Hisano, K.~Ishiwata and N.~Nagata,
  Phys.\ Rev.\ D {\bf 82}, 115007  (2010).

\bibitem{Hisano:2011cs} 
  J.~Hisano, K.~Ishiwata, N.~Nagata and T.~Takesako,
  JHEP {\bf 1107}, 005 (2011)
.

\bibitem{Peccei:1977hh} 
  R.~D.~Peccei and H.~R.~Quinn,
  Phys.\ Rev.\ Lett.\  {\bf 38}, 1440 (1977).

\bibitem{Drees:1993bu}
  M.~Drees and M.~Nojiri,
  Phys.\ Rev.\  D {\bf 48}, 3483 (1993).

\bibitem{Jungman:1995df}
G.~Jungman, M.~Kamionkowski and K.~Griest,
Phys.\ Rept.\  {\bf 267}, 195 (1996).



\bibitem{Djouadi2000}
A. Djouadi and M. Drees, Phys.\ Lett.\ B {\bf 484}, 183 (2000).

\bibitem{Hill:2011be} 
  R.~J.~Hill and M.~P.~Solon,
  Phys.\ Lett.\ B {\bf 707}, 539 (2012)
.

\bibitem{Young:2009zb} 
  R.~D.~Young and A.~W.~Thomas,
  Phys.\ Rev.\ D {\bf 81}, 014503 (2010)
.

\bibitem{:2012sa} 
 H. Ohki , {\it et al.}  [JLQCD Collaboration],
  arXiv:1208.4185 [hep-lat].

\bibitem{Pumplin:2002vw}
J.~Pumplin, D.~R.~Stump, J.~Huston, H.~L.~Lai, P.~Nadolsky and W.~K.~Tung,
JHEP {\bf 0207}, 012  (2002) .

\bibitem{Adams:1995ufa}
  D.~Adams {\it et al.}  [Spin Muon Collaboration],
  Phys.\ Lett.\  B {\bf 357},  248  (1995).

\bibitem{Hisano:2004pv} 
  J.~Hisano, S.~Matsumoto, M.~M.~Nojiri and O.~Saito,
  Phys.\ Rev.\ D {\bf 71}, 015007 (2005).

\bibitem{Aprile:2012nq} 
  E.~Aprile {\it et al.}  [XENON100 Collaboration],
  arXiv:1207.5988 [astro-ph.CO].

\bibitem{Gutlein:2010tq} 
  A.~Gutlein, C.~Ciemniak, F.~von Feilitzsch,
  N.~Haag, M.~Hofmann, C.~Isaila, T.~Lachenmaier and J.~-C.~Lanfranchi
  {\it et al.},
  Astropart.\ Phys.\  {\bf 34}, 90 (2010).

\bibitem{Pavan:2001wz} 
  M.~M.~Pavan, I.~I.~Strakovsky, R.~L.~Workman and R.~A.~Arndt,
  PiN Newslett.\  {\bf 16}, 110 (2002)~.

\bibitem{Borasoy:1996bx} 
  B.~Borasoy and U.~-G.~Meissner,
  Annals Phys.\  {\bf 254}, 192 (1997)~.

\bibitem{Alarcon:2012nr} 
  J.~M.~Alarcon, L.~S.~Geng, J.~M.~Camalich and J.~A.~Oller,
  arXiv:1209.2870 [hep-ph].

\end{thebibliography}
\end{document}